\documentclass[aps,pra,10pt,twocolumn,superscriptaddress]{revtex4-2}

\usepackage{amsmath, amsfonts}
\usepackage{comment}
\usepackage{graphicx}
\usepackage{braket}
\usepackage{qcircuit}
\usepackage{xcolor}
\usepackage{amssymb}
\usepackage{hyperref}
\usepackage{cleveref}
\usepackage{array}

\usepackage[caption=false]{subfig}
\usepackage{float}

\newcommand{\id}{I}

\begin{document}
\preprint{APS/123-QED}
\title{
Demonstration~of~a~Logical~Architecture~Uniting~Motion~and~In-Place~Entanglement
}
\author{Rich Rines}
\thanks{These authors contributed equally to this work.}
\affiliation{Infleqtion (Boulder, CO, USA; Chicago, IL, USA; Louisville, CO, USA; Madison, WI, USA; Oxford, England, UK)}

\author{Benjamin Hall}
\thanks{These authors contributed equally to this work.}
\affiliation{Infleqtion (Boulder, CO, USA; Chicago, IL, USA; Louisville, CO, USA; Madison, WI, USA; Oxford, England, UK)}

\author{Mariesa H. Teo}
\thanks{These authors contributed equally to this work.}
\affiliation{Pritzker School of Molecular Engineering, University of Chicago, Chicago, IL, USA}

\author{Joshua Viszlai}
\thanks{These authors contributed equally to this work.}
\affiliation{Infleqtion (Boulder, CO, USA; Chicago, IL, USA; Louisville, CO, USA; Madison, WI, USA; Oxford, England, UK)}
\affiliation{Department of Computer Science, University of Chicago, Chicago, IL, USA}

\author{Daniel C. Cole}
\thanks{These authors contributed equally to this work.}
\affiliation{Infleqtion (Boulder, CO, USA; Chicago, IL, USA; Louisville, CO, USA; Madison, WI, USA; Oxford, England, UK)}

\author{David Mason}
\thanks{These authors contributed equally to this work.}
\affiliation{Infleqtion (Boulder, CO, USA; Chicago, IL, USA; Louisville, CO, USA; Madison, WI, USA; Oxford, England, UK)}

\author{Cameron Barker}
\affiliation{Infleqtion (Boulder, CO, USA; Chicago, IL, USA; Louisville, CO, USA; Madison, WI, USA; Oxford, England, UK)}

\author{Matt J. Bedalov}
\affiliation{Infleqtion (Boulder, CO, USA; Chicago, IL, USA; Louisville, CO, USA; Madison, WI, USA; Oxford, England, UK)}

\author{Matt Blakely}
\affiliation{Infleqtion (Boulder, CO, USA; Chicago, IL, USA; Louisville, CO, USA; Madison, WI, USA; Oxford, England, UK)}

\author{Tobias Bothwell}
\affiliation{Infleqtion (Boulder, CO, USA; Chicago, IL, USA; Louisville, CO, USA; Madison, WI, USA; Oxford, England, UK)}

\author{Caitlin Carnahan}
\affiliation{Infleqtion (Boulder, CO, USA; Chicago, IL, USA; Louisville, CO, USA; Madison, WI, USA; Oxford, England, UK)}

\author{Frederic T. Chong}
\affiliation{Infleqtion (Boulder, CO, USA; Chicago, IL, USA; Louisville, CO, USA; Madison, WI, USA; Oxford, England, UK)}
\affiliation{Department of Computer Science, University of Chicago, Chicago, IL, USA}

\author{Samuel Y. Eubanks}
\affiliation{Infleqtion (Boulder, CO, USA; Chicago, IL, USA; Louisville, CO, USA; Madison, WI, USA; Oxford, England, UK)}

\author{Brian Fields}
\affiliation{Infleqtion (Boulder, CO, USA; Chicago, IL, USA; Louisville, CO, USA; Madison, WI, USA; Oxford, England, UK)}

\author{Matthew Gillette}
\affiliation{Infleqtion (Boulder, CO, USA; Chicago, IL, USA; Louisville, CO, USA; Madison, WI, USA; Oxford, England, UK)}

\author{Palash Goiporia}
\affiliation{Infleqtion (Boulder, CO, USA; Chicago, IL, USA; Louisville, CO, USA; Madison, WI, USA; Oxford, England, UK)}

\author{Pranav Gokhale}
\affiliation{Infleqtion (Boulder, CO, USA; Chicago, IL, USA; Louisville, CO, USA; Madison, WI, USA; Oxford, England, UK)}

\author{Garrett T. Hickman}
\affiliation{Infleqtion (Boulder, CO, USA; Chicago, IL, USA; Louisville, CO, USA; Madison, WI, USA; Oxford, England, UK)}

\author{Marin Iliev}
\affiliation{Infleqtion (Boulder, CO, USA; Chicago, IL, USA; Louisville, CO, USA; Madison, WI, USA; Oxford, England, UK)}

\author{Eric B. Jones}
\affiliation{Infleqtion (Boulder, CO, USA; Chicago, IL, USA; Louisville, CO, USA; Madison, WI, USA; Oxford, England, UK)}

\author{Ryan A. Jones}
\affiliation{Infleqtion (Boulder, CO, USA; Chicago, IL, USA; Louisville, CO, USA; Madison, WI, USA; Oxford, England, UK)}

\author{Kevin W. Kuper}
\affiliation{Infleqtion (Boulder, CO, USA; Chicago, IL, USA; Louisville, CO, USA; Madison, WI, USA; Oxford, England, UK)}

\author{Stephanie Lee}
\affiliation{Infleqtion (Boulder, CO, USA; Chicago, IL, USA; Louisville, CO, USA; Madison, WI, USA; Oxford, England, UK)}

\author{Martin T. Lichtman}
\affiliation{Infleqtion (Boulder, CO, USA; Chicago, IL, USA; Louisville, CO, USA; Madison, WI, USA; Oxford, England, UK)}

\author{Kevin Loeffler}
\affiliation{Infleqtion (Boulder, CO, USA; Chicago, IL, USA; Louisville, CO, USA; Madison, WI, USA; Oxford, England, UK)}

\author{Nate Mackintosh}
\affiliation{Infleqtion (Boulder, CO, USA; Chicago, IL, USA; Louisville, CO, USA; Madison, WI, USA; Oxford, England, UK)}

\author{Farhad Majdeteimouri}
\affiliation{Infleqtion (Boulder, CO, USA; Chicago, IL, USA; Louisville, CO, USA; Madison, WI, USA; Oxford, England, UK)}

\author{Peter T. Mitchell}
\affiliation{Infleqtion (Boulder, CO, USA; Chicago, IL, USA; Louisville, CO, USA; Madison, WI, USA; Oxford, England, UK)}

\author{Thomas W. Noel}
\affiliation{Infleqtion (Boulder, CO, USA; Chicago, IL, USA; Louisville, CO, USA; Madison, WI, USA; Oxford, England, UK)}

\author{Ely Novakoski}
\affiliation{Infleqtion (Boulder, CO, USA; Chicago, IL, USA; Louisville, CO, USA; Madison, WI, USA; Oxford, England, UK)}

\author{Victory Omole}
\email{victory.omole@infleqtion}
\affiliation{Infleqtion (Boulder, CO, USA; Chicago, IL, USA; Louisville, CO, USA; Madison, WI, USA; Oxford, England, UK)}

\author{David Owusu-Antwi}
\affiliation{Infleqtion (Boulder, CO, USA; Chicago, IL, USA; Louisville, CO, USA; Madison, WI, USA; Oxford, England, UK)}

\author{Alexander G. Radnaev}
\affiliation{Infleqtion (Boulder, CO, USA; Chicago, IL, USA; Louisville, CO, USA; Madison, WI, USA; Oxford, England, UK)}

\author{Anthony Reiter}
\affiliation{Infleqtion (Boulder, CO, USA; Chicago, IL, USA; Louisville, CO, USA; Madison, WI, USA; Oxford, England, UK)}

\author{Mark Saffman}
\affiliation{Infleqtion (Boulder, CO, USA; Chicago, IL, USA; Louisville, CO, USA; Madison, WI, USA; Oxford, England, UK)}
\affiliation{Department of Physics, University of Wisconsin–Madison, Madison, WI, USA}

\author{Bharath Thotakura}
\affiliation{Infleqtion (Boulder, CO, USA; Chicago, IL, USA; Louisville, CO, USA; Madison, WI, USA; Oxford, England, UK)}

\author{Teague Tomesh}
\affiliation{Infleqtion (Boulder, CO, USA; Chicago, IL, USA; Louisville, CO, USA; Madison, WI, USA; Oxford, England, UK)}

\author{Ilya Vinogradov}
\affiliation{Infleqtion (Boulder, CO, USA; Chicago, IL, USA; Louisville, CO, USA; Madison, WI, USA; Oxford, England, UK)}
\date{\today}
\begin{abstract}
We demonstrate a logical neutral atom architecture that integrates atom motion with in‑place entanglement to achieve lower overheads than entangling‑zone approaches. Using a 114‑qubit device, we perform three proof‑of‑principle logical‑qubit experiments. First, we implement a pre‑compiled, non‑scalable variant of Shor’s algorithm, observing improved logical‑over‑physical performance, including with loss correction and leakage detection, achieving up to a ~2× reduction in TVD. Second, we construct constant‑depth logical CX ladders; on current hardware these execute with serial entangling operations, yet still yield ~2-4× lower error for 8 and 12 logical qubits. Third, we prepare the [[16,4,4]] code and perform single‑round decoding with post‑processed error correction, achieving ~8× improvement on logical vs physical. These results demonstrate how combining motion with in‑place entanglement offers lower overhead than entangling‑zone approaches.
\end{abstract}
\maketitle

\section{Introduction}
In the past few years, multiple experiments have achieved control of logical qubits \cite{bluvstein2022, google2023suppressing, paetznick2024demonstration, bedalov2024fault, rodriguez2025experimentaldemonstrationlogicalmagic, reichardt2024demonstration, bluvstein2025logical}. Most of the experiments leverage non-local qubit connectivity to achieve significant milestones such as high-rate quantum error correction codes, efficient magic state preparation, universal gatesets, and code switching. The demonstrations that leverage non-local connectivity typically perform two-qubit physical operations with an architecture that moves qubits into an entanglement zone. Although the zone-based architecture lowers the requirements for efficient classical control, gate calibration, and parallelism\cite{bluvstein2024logical}, it is not yet clear if it will satisfy the threshold theorem for large-scale fault-tolerant quantum computation \cite{sunami2025}. In this paper, we realize an alternative neutral atom architecture that unites qubit motion with individual optical addressing for in-place entangling operations. This approach was demonstrated recently on a small scale with physical qubits \cite{Chinnarasu2025} and is here extended to logical qubit (LQ) encodings. The architecture maintains all-to-all connectivity, while reducing the pitfalls associated with atom motion like runtime cost and error accumulation.

\begin{figure}[h!]
    \centering
    \includegraphics[width=\linewidth]{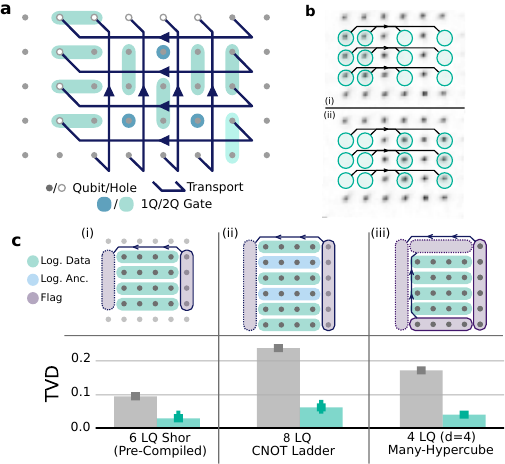}
    \caption{
    (a) Sqale's architecture, combining locally addressed gates and mid-circuit rearrangement. (b) Fluorescence images before (top) and after (bottom) translating and stretching a 6-qubit block through interstitial lanes. (c) Summary of demonstrations in this work.  Upper panel: logical qubit layout and mid-circuit reconfigurations for (i, ii) [[4, 2, 2]]-encoded demonstration of Shor's algorithm and a constant-depth CNOT (CX) ladder, and (iii) state preparation in the [[16, 4, 4]] many-hypercube code.  Lower panel: Encoded/unencoded performance in terms of total variation distance (TVD).
    }
    \label{fig:summary_of_results}
\end{figure}

We demonstrate this architecture by experimentally executing multiple circuits on Infleqtion's Sqale QPU that show primitives on the path to utility-scale quantum computing. The remainder of this paper is structured as follows:
\begin{itemize}
    \item Section \ref{sec:methods} describes the methods underpinning the Sqale architecture from a software \& hardware perspective.
    \item Section \ref{sec:shor} describes the pre-compiled Shor’s Algorithm circuits \cite{smolin2013oversimplifyingquantumfactoring} executed with logical qubits on hardware, including with loss correction and with leakage detection. This constitutes the first demonstration of Shor's Algorithm on logical qubits, although we emphasize that the pre-compiled circuits are unscalable.
    \item Section \ref{sec:cdcx} describes a novel technique for implementing logical constant-depth CX ladders, with no depth dependence on either the number of logical qubits $N$ or the code distance $d$ of the system. The CX ladder circuits leverage ancilla logical qubits and transversal gates. We show proof-of-principle experiments with up to 12 logical qubits.
    \item Section~\ref{sec:hypercube} describes the execution of state preparation under the level-2 $[[4^{L}, 2^{L}, 2^{L}]]$ many-hypercube code \cite{goto2024high}, which forms a [[16, 4, 4]] code.
    \item Section~\ref{sec:outlook} presents a broad reflection on how the experiments that were performed herein contribute to the architectural principles for neutral atom quantum computing. Section~\ref{sec:conclusion} concludes.
\end{itemize}

These experiments serve as stepping stones towards running utility-scale applications on neutral atom quantum computers, such as elucidating the properties of optoelectronic materials \cite{jones2024dynamic}. For instance, the pre-compiled Shor's Algorithm circuit is a Hadamard test, a crucial primitive for some strongly correlated electron models, such as the Anderson impurity model \cite{jones2024dynamic}. In turn, the Trotterized implementation of the Anderson impurity Hamiltonian will benefit from constant-depth CX ladders \cite{Baumer2025}. Moreover, the relatively slow physical gate times, combined with the capacity to trap large numbers of qubits \cite{manetsch2025tweezerarray6100highly, chiu2025continuousoperationofacoherent3} make neutral atom systems well-suited for constant-depth circuits\cite{litinski2019gameofsurfacecodes}. Finally, the many-hypercube code is a contender for low-overhead utility-scale quantum computation \cite{litinski2025fault}.
\section{Methods}\label{sec:methods}
\subsection{Movement}

In this work, we demonstrate the added capability of mid-circuit atom rearrangement on top of a neutral atom platform with individual optical addressability and non-destructive qubit readout \cite{radnaev2024universal}. Mid-circuit atom rearrangement enables all-to-all connectivity \cite{bluvstein2024logical}. In the context of our demonstrations, this capability enables fault-tolerant initialization with flag qubits, transversal entangling logic, and circuit-based leakage detection that would otherwise be challenging with typical planar connectivity.
Specifically, our experiments leverage block (parallel) atom motion to achieve the necessary connectivities---typically ring or torus. Once atom motion is performed, several CZ gates are subsequently performed in-place, without necessitating motion to an entangling zone. As such, our experiments exemplify the potential of our underlying architecture, which uses motion sparingly.

Our experiments are executed on a 7 $\times$ 6 = 42 qubit patch of 114 qubits within Sqale's array of 14 $\times$ 16 = 224 optical tweezers. The array has spacing of $6~\mu\mathrm{m}$, allowing qubits to be moved between sites. Multiple atoms may be moved simultaneously. Further information about the mid-circuit rearrangement capability, as well as single- and two-qubit gate performance, is provided in Section \ref{sec:app_hardware}.

\subsection{Logical Encoding}
\label{subsection:logical_encoding}
The realization of high-performance quantum error-correcting (QEC) codes is essential for enabling fault-tolerant quantum computation (FTQC) at scale. Physical qubits are intrinsically fragile, as they are subject to decoherence, gate imperfections, and environmental noise. To mitigate these effects, imperfect physical qubits can be redundantly encoded into logical qubits that achieve significantly higher gate fidelities. The characteristics of QEC codes---such as their encoding rates, resource overhead, and connectivity requirements---vary across code families, making certain codes more suitable than others for specific hardware architectures. In this work, we encode our logical qubits in 2 different codes: the [[4, 2, 2]] code and the [[16, 4, 4]] many-hypercube code.  Each circuit is compiled to Sqale's native gateset \cite{radnaev2024universal}, which includes state preparation, two-qubit CZ gates, single-qubit global rotation $GR_{\theta, \phi}$ and $R_z$ gates, atom movement, and measurement.\\

The [[4, 2, 2]] ($C_4$) code is a Calderbank-Shor-Steane (CSS) stabilizer code that encodes 2 logical qubits within 4 physical qubits. It has a distance of $d=2$, meaning that a single error can be detected but not corrected. Its stabilizers are $XXXX$ and $ZZZZ$, admitting the following encoding:
\begin{align}
\ket{00}_L &= \left(\ket{0000} + \ket{1111}\right) / \sqrt{2} \nonumber \\
\ket{01}_L &= \left(\ket{0011} + \ket{1100}\right) / \sqrt{2} \nonumber \\
\ket{10}_L &= \left(\ket{0101} + \ket{1010}\right) / \sqrt{2} \nonumber \\
\ket{11}_L &= \left(\ket{0110} + \ket{1001}\right) / \sqrt{2}
\label{eq:codewords}
\end{align}

For this code, the CX gate is transversal. The encodings for the fault-tolerant logical gateset, as well as their corresponding compilations to Sqale's native gateset, can be found in the appendix of our previous work \cite{bedalov2024fault}. Compilations for the circuits were performed with the Superstaq \cite{campbell2023superstaq} optimizing compiler.

In our previous work \cite{bedalov2024fault}, the $[[4,2,2]]$ code was implemented within a static hexagonal lattice that permitted the ring connectivity required for fault-tolerant encodings. In this paper the qubits are arranged in a square grid, and we use motion to achieve ring connectivity.

While the [[4, 2, 2]] code is an error detection code, recent work has shown that it can be concatenated with other detection codes to produce high threshold error correction codes with encoding rates 10-100x better than the surface code \cite{yoshida2025concatenate}. To that end, in Section \ref{sec:hypercube} we demonstrate preparation of a [[16, 4, 4]] many-hypercube code patch \cite{goto2024high} as a level-2 concatenation of the [[4, 2, 2]] code with itself.

\subsection{Post-Processing}\label{subsection:post-processing}
During circuit execution, measurements are collected from hardware and then post-processed via (atom) loss detection, (atom) loss correction, decoding, and/or Pauli correction. In general, we postselect on atom loss, except when performing \textit{loss correction}. None of the post-processing methods that we apply are scaleable.

\textbf{Loss detection}: A result is discarded whenever an atom is not detected at an expected site during measurement. 

\textbf{Loss correction}: As an alternative to postselection on atom loss, we can instead attempt to \textit{correct} for single loss events by inferring a state for the lost atom that would complete a logical codeword. For instance, upon reading out $\ket{111L}$, where $\ket{L}$ indicates an atom loss, the decoder can correct to $\ket{1111}$ which corresponds to $\ket{00}_L$. However, the increased acceptance rate comes at the expense of decoder confidence (and therefore logical fidelity): a single additional bit flip will cause us to miscorrect the lost qubit and infer the wrong logical state. We leave implementation of more sophisticated loss correction decoding that is aware of underlying physical circuits \cite{stricker2020experimental, reichardt2024fault, baranes2025leveraging} to future work.

\textbf{Decoding}: To know if a measurement can be mapped to a logical state, it must first be checked for non-zero flags and out-of-codespace patches. 
Non-zero measurements of flag qubits indicate that its corresponding patch of data qubits was not correctly initialized into the logical $|00\rangle_L$ state. 
Any measurement for which not all flags are measured to be zero is discarded. A patch of data qubits is out-of-codespace if it does not match one of the codewords of the [[4, 2, 2]] code (Equation \ref{eq:codewords}). Any measurement for which not all patches are within the codespace is discarded. Once these checks are completed, the remaining measurements are decoded via Equation \ref{eq:codewords}.

\textbf{Pauli Correction (for constant depth circuits)}: Appropriate Pauli corrections (bit flips) are applied to the data qubit measurements based on the ancilla qubit measurements.
This step is the result of pushing the classically controlled Pauli corrections (required to make a circuit constant-depth) past the measurements. 

\subsection{Leakage Detection}

Non-loss leakage errors occur when the state of a physical qubit escapes the two-level subspace spanned by the prescribed computational basis states $\ket0$ and $\ket1$. On Sqale, non-loss leakage predominantly occupies the $m_F\neq0$ Zeeman sublevels of the $F=3$ and $F=4$ manifolds of the Cs ground state, which are respectively indistinguishable from the $\ket0$ and $\ket1$ computational states during readout.
Although non-loss leakage can be converted to loss at the hardware level \cite{reichardt2024fault, zhang2025leveragingerasureerrorslogical}, leakage detection units (LDUs) detect non-loss leakage without hardware modifications \cite{Preskill1998b,chow2024}.
Logically, each LDU comprises two CXs targeting a flag qubit, conditioned on opposite states of the qubit to be observed:
\begin{equation}
\Qcircuit @R=0.1em @C=0.3em @!R {
 &\lstick{}       \qw & \qw & \ctrl{1} & \qw  & \ctrlo{1} & \qw & \qw \\
 &\lstick{\ket{1}}    & \qw & \targ    & \qw  & \targ     & \qw & \meter \\
}
\end{equation}
The flag will be $\ket0$ if and only if the control qubit is in the qubit subspace because the CX gates have no effect when the control is in a non-loss leaked state.

\subsection{Noisy Simulations}
Quantum circuit simulation under realistic noise modeling is fundamental to the design and validation of error-corrected quantum computation, and relies heavily on modeling errors specific to the qubit modality of interest \cite{cong2022hardware, sahay2025foldtransversalsurfacecodecultivation}.
Our Sqalesim simulator \cite{bedalov2024fault} models each atom as a five-level system composed of the computational basis qubit subspace $\{|0\rangle, |1\rangle \}$, two leakage states unaffected by gates and detected as $|0\rangle$ and $|1\rangle$, and a loss state. Whereas the previous implementation of Sqalesim \cite{bedalov2024fault} relied primarily on density matrix simulation, the current implementation achieves greater efficiency by simulating operations that only affect the qubit subspace through a Clifford-only simulator. Any non-Clifford operation is replaced with its Pauli Twirling approximation. Transitions in and out of the qubit subspace are tracked using a classical distribution that describes the probability of the atom being in each non-computational state. The current implementation includes a simple model of atom motion in which each atom being moved has a probability that its qubit subspace undergoes a phase error.
\section{Shor's Algorithm (Pre-Compiled)}\label{sec:shor}

The heart of Shor's Algorithm \cite{shor1994algorithms} is the quantum order-finding subroutine, which determines the period $r$ such that $a^r\equiv 1\pmod{N}$ for a given base $a$ and semiprime integer $N$ to be factored. The order-finding algorithm in turn comprises the following circuit:
\begin{equation*}
\centering
\label{eq:shor:order_finding}
\Qcircuit @C=1.2em @R=.9em{
\lstick{|+\rangle} & \qw / & \qw & \ctrl{1} & \gate{\mathrm{QFT}^{\dagger}} & \meter & \cw \\
\lstick{|1\rangle} & \qw / & \qw & \gate{a^{\,x}\ \bmod\ N}               & \qw                       & \qw    & \qw
}
\end{equation*}
In this circuit (where / denotes a bundle of quantum wires), the top register is initialized to the uniform superposition $|+\rangle$ state and acts as the control for the modular exponentiation. The bottom register is initialized to $|1\rangle$ and undergoes the modular multiplication operation $a^{x}\mod N$, where the exponent $x$ is determined by the state of the control register. Following modular exponentiation, the inverse Quantum Fourier Transform (QFT) is applied to the control register, which enables extraction of the period, $r$, after final measurement. Information about $r$ is then classically processed via the continued fractions algorithm, allowing deduction of the factors of $N$.
    
In general, the base $a$ is to be chosen randomly. However, there exist choices such that $a^{2^{k}}\equiv1\pmod{N}$ for some $k$, causing the sequence to terminate and the circuit to be smaller. It is known that such an $a$ always exists for $k=1$ \cite{smolin2013oversimplifyingquantumfactoring}, at which point the circuit reduces to a Hadamard test with a single controlled-modular multiplier. The multiplier itself has a known input and can therefore be replaced with a few CXs. 
Such pre-compiled constructions are inherently unscalable because they depend on prior knowledge that is effectively equivalent to knowing the answer in advance. A truly scalable version of Shor’s Algorithm instead coherently executes the modular arithmetic on the quantum computer itself, without access to this hidden information.
As a result, the complexity of the order-finding algorithm can be significantly reduced with special values of the base $a$ for modular exponentiation. For $N=15$, the choices are $a\in\{4, 11, 14\}$. Taking $a=11$ for $N=15$, the pre-compiled order-finding circuit is simply:
\begin{equation}
\label{eq:shor:logical}
\Qcircuit @R=0.1em @C=0.3em @!R {
 &\lstick{\ket{0}}& \qw & \gate{H} & \qw & \ctrl{1} & \qw & \ctrl{2} & \qw & \gate{H} & \qw & \meter\\
 &\lstick{\ket{0}}& \qw & \qw      & \qw & \targ    & \qw & \qw      & \qw & \qw      & \qw & \qw \\
 &\lstick{\ket{0}}& \qw & \qw      & \qw & \qw      & \qw & \targ    & \qw & \qw      & \qw & \qw \\
 &\lstick{\ket{0}}& \qw & \qw      & \qw & \qw      & \qw & \qw      & \qw & \qw      & \qw & \qw \\
 &\lstick{\ket{1}}& \qw & \qw      & \qw & \qw      & \qw & \qw      & \qw & \qw      & \qw & \qw \\
}.
\end{equation}
\newline
For our experiments we drop the two unused qubits, leaving just the three-qubit Hadamard test circuit. We refer to the Sqale compiled version of this as the `unencoded Shor' circuit.

Though the order-finding algorithm only requires measurements from the top (QFT register) qubit, we validate our experiments by measuring all three non-idle qubits and compare the resulting distribution to that of the ideal distribution using the metric of Total Variation Distance (TVD).

\subsection{Logical circuits}

We execute two different embeddings of the unencoded Shor circuit into the [[4, 2, 2]] code. These embeddings use the code's fault tolerant state preparation and transversal operations.
A two-row embedding requires two logical patches, but makes use of three of the four resulting logical qubits:
\begin{equation}
\label{eq:shor:2row}
\Qcircuit @R=0.5em @C=0.4em {
 \lstick{}& \qw & \gate{H} & \qw & \qswap     & \qw & \qw & \ctrl{1}    & \qw & \qw & \ctrl{3} & \qw      & \qw & \qw & \gate{H} & \qw & \qswap     & \qw & \qw & \meter \\
 \lstick{}& \qw & \gate{H} & \qw & \qswap\qwx & \qw & \qw & \control\qw & \qw & \qw & \qw      & \ctrl{3} & \qw & \qw & \gate{H} & \qw & \qswap\qwx & \qw & \qw & \meter
 \inputgroupv{1}{2}{0.8em}{0.8em}{\ket{00}_L\;\;}\\
 &\\
 \lstick{}& \qw & \qw      & \qw & \qw        & \qw & \qw & \qw         & \qw & \qw & \targ    & \qw      & \qw & \qw & \qw      & \qw & \qw        & \qw & \qw & \meter \\
 \lstick{}& \qw & \qw      & \qw & \qw        & \qw & \qw & \qw         & \qw & \qw & \qw      & \targ    & \qw & \qw & \qw      & \qw & \qw        & \qw & \qw & \qw
 \inputgroupv{4}{5}{0.8em}{0.8em}{\ket{0+}_L\;\;}\\ 
}
\end{equation}
The two registers are arranged linearly in adjacent rows, ensuring that the transversal CX gates are all performed between neighboring qubits.
The lower patch is initialized in the $\ket{0+}_L$ state, so the CX gate has no effect on the (unused) $\ket{+}$ qubit. The remaining operations are then equivalent to those in Equation \ref{eq:shor:logical} on the three $\ket0$-initialized logical qubits.
The preparation of $\ket{00}_L$ in the upper register requires a flag qubit and ring connectivity, as described in Section \ref{sec:methods}.
We achieve the necessary connectivity via a single atom move, transporting the flag qubit from one end of the patch to the other to allow CXs from the data qubits on either side.
The circuit is compiled to the Sqale native gateset, and the final compiled circuit contains 11 CZ gates, 5 GR gates, and one movement operation.

The two-row Shor experiment is also repeated with LDUs on the eight data qubits near the end of the circuit before the final round of Hadamards in Equation \ref{eq:shor:2row}. This allows for a more efficient compilation than if they were placed at the end of the circuits. Operations that follow the LDUs are not expected to contribute substantially to leakage errors.
In post-processing, a nonzero measurement of any LDU flag qubit is treated identically to an observed atom loss on the corresponding data qubit.
These flag qubits are placed in additional rows of atoms above and below the two logical patches. In total, the addition of LDUs requires an additional 16 CZ gates and one GR.

A three-row embedding uses three logical patches, but avoids the unused logical qubit in the two-patch variant. Instead, the three-qubit circuit is implemented transversally across the three registers, so that in effect it is replicated for each logical qubit in each register:
\begin{equation}
\Qcircuit @R=0.5em @C=0.4em {
 \lstick{}& \qw & \qw      & \qw & \qw        & \qw & \qw & \targ     & \qw       & \qw      & \qw      & \qw & \qw & \qw      & \qw & \qw        & \qw & \qw & \meter \\
 \lstick{}& \qw & \qw      & \qw & \qw        & \qw & \qw & \qw       & \targ     & \qw      & \qw      & \qw & \qw & \qw      & \qw & \qw        & \qw & \qw & \meter
 \inputgroupv{1}{2}{0.8em}{0.8em}{\ket{00}_L\;\;}\\        
 &\\                                                       
 \lstick{}& \qw & \gate{H} & \qw & \qswap     & \qw & \qw & \ctrl{-3} & \qw       & \ctrl{3} & \qw      & \qw & \qw & \gate{H} & \qw & \qswap     & \qw & \qw & \meter \\
 \lstick{}& \qw & \gate{H} & \qw & \qswap\qwx & \qw & \qw & \qw       & \ctrl{-3} & \qw      & \ctrl{3} & \qw & \qw & \gate{H} & \qw & \qswap\qwx & \qw & \qw & \meter
 \inputgroupv{4}{5}{0.8em}{0.8em}{\ket{00}_L\;\;}\\        
 &\\                                                       
 \lstick{}& \qw & \qw      & \qw & \qw        & \qw & \qw & \qw       & \qw       & \targ    & \qw      & \qw & \qw & \qw      & \qw & \qw        & \qw & \qw & \meter \\
 \lstick{}& \qw & \qw      & \qw & \qw        & \qw & \qw & \qw       & \qw       & \qw      & \targ    & \qw & \qw & \qw      & \qw & \qw        & \qw & \qw & \meter
 \inputgroupv{7}{8}{0.8em}{0.8em}{\ket{00}_L\;\;}\\
}
\end{equation}
\newline
As a result, each run of the encoded circuit generates two independent samples of the circuit.

Like the two-row embedding, logical patches are stacked on adjacent rows of atoms to allow transversal CXs between adjacent patches. In this case all three patches are initialized to $\ket{00}_L$, requiring their own flag qubits. The three flags are initially placed in a single column on one side of the logical patches, and are simultaneously transported to the other end of their respective registers in a single motion operation.

The three-row embedding requires twice as many CZs (22) and the same number of GRs (5) as the two-row embedding when compiled to the Sqale gateset.

\subsection{Results}

\begin{figure}
    a.\\
    \includegraphics[width=.92\linewidth]{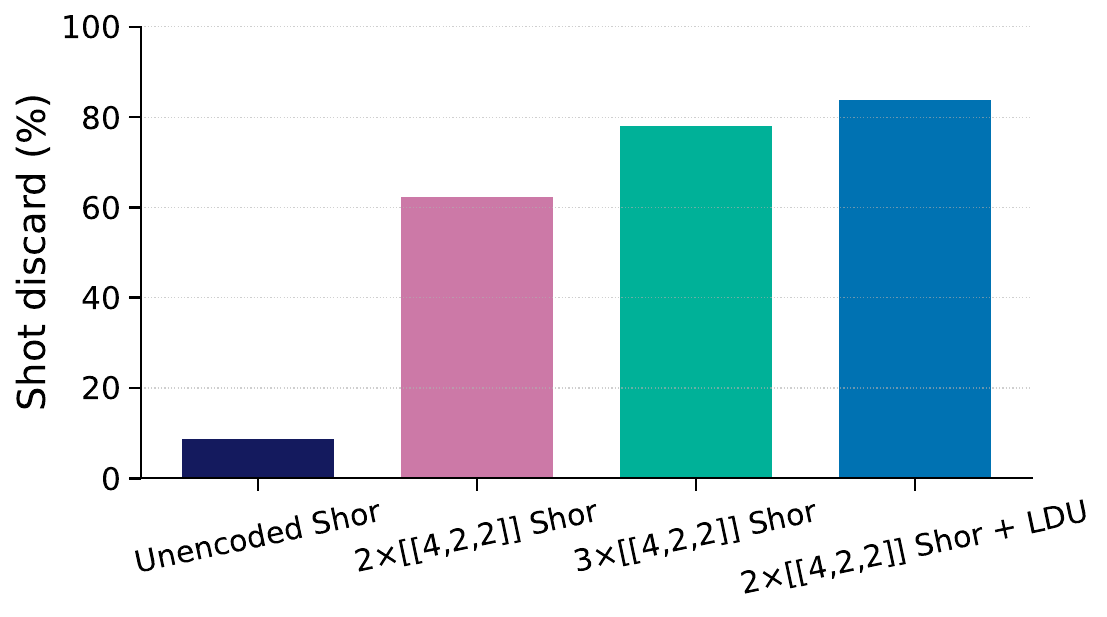}\\
    b.\\
    \includegraphics[width=.94\linewidth]{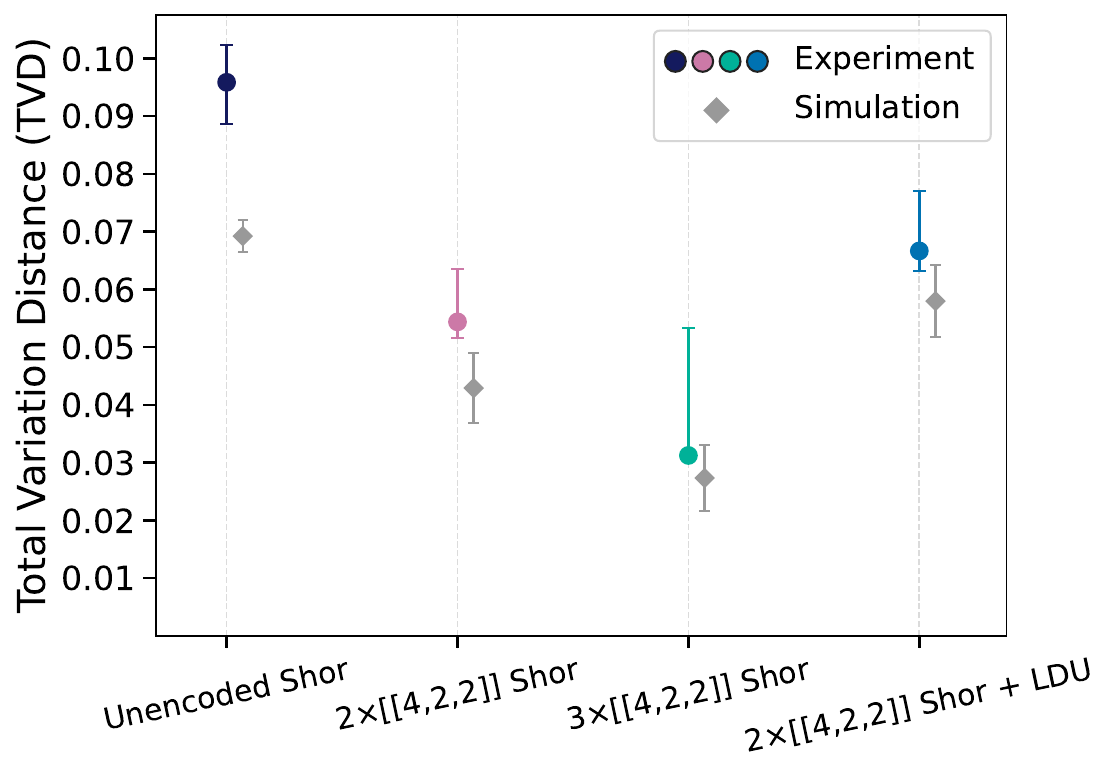}
    \caption{Experimental results for all Shor circuits run on hardware.
    (a) The postselection discard, and
    (b) The TVD computed between the experimental and ideal distributions in (a). The error bars represent a 68\% confidence interval (CI) on the TVD for the narrowest CI obtained through minimization of the CI width. Diamond data points represent the expected (mean) TVD from a default Sqalesim noise sampling for the same number of experimental shots, with error bars on the mean of ten trials.}
    \label{fig:shor_experimental_data}
\end{figure}

Experimental results from four implementations: unencoded, two-row embedding, three-row logical embedding, and two-row with leakage detection, as well as those predicted via noisy simulation, are displayed in Fig. \ref{fig:shor_experimental_data}. All results are postselected on atom loss (see Section \ref{sec:loss correction} for results with loss correction post-processing instead). The three encoded implementations are additionally postselected on flag qubits for their $\ket{00}_L$ preparation and the parity of the measured data qubits, as well as (in the case of the leakage-detecting circuit) the measured state of the eight LDU flags.
At least 5$\times10^{3}$ shots were taken per circuit, with $10^{4}$ shots of the leakage detection circuit to compensate for its low postselection yield.

We see statistically significant reduction in the TVDs of the encoded circuits relative to the unencoded circuit.
Notably, despite getting two samples per shot, we find that after the same number of shots, the three-row logical circuit has substantially fewer valid samples after postselection than the two-row variant (1152 vs. 1986). At the same time, we see a discernible drop in TVD between the two-row embedding to the three-row embedding, more so than initially predicted by their noisy simulation (investigated further in Section \ref{sec:shor_noisy_simulations}). 
Though they resulted in a substantially lower postselection yield, we did not see an advantage (in terms of TVD) from using the leakage detection units (LDU). We attribute this to the considerable noise added from the additional two-qubit gates, and expect that the same units would prove beneficial within a deeper circuit in which the leakage is more likely to have occurred.
Indeed, we do see a substantial increase in TVD (from 6.7\% shown to 14.1\%) if we reinterpret the same results without postselecting on the LDU flags.

\section{Constant-Depth CX ladders}\label{sec:cdcx}

A CX ladder is a series of CX (controlled-NOT) gates used to compute the binary sum (mod 2) of a set of qubits' states \cite{Baumer2025}. CX ladders are a crucial sub-component of many fundamental circuits in quantum computing, including modular exponentiation for Shor's algorithm \cite{shor1994algorithms}, GHZ-state preparation, and implementation of Pauli phasors for Hamiltonian simulation.

Because addition is both commutative and associative, the CX ladder may be arranged in any manner that achieves this summation.
As we wish to ultimately encode a CX ladder via the [[4, 2, 2]] code on a nearest-neighbor architecture, we consider the ``outside-in'' CX ladder arrangement, a 5-CX example of which is shown in Figure \ref{fig:cx_half}. It only requires linear nearest-neighbor connectivity and it can be embedded into a $2\times3$ grid. This is shown via the double-indexed qubit labels after the arrows.

\begin{figure}[h]
    \centering
    \subfloat[]{
    \centering
        \Qcircuit @C=3.5em @R=1em {
        \lstick{d_{1} \to \textcolor{cyan}{d_{1,0}}}   & \ctrl{1} \barrier[-2em]{5} & \qw       \barrier[-2em]{5} & \qw      & \meter \\ 
        \lstick{d_{2} \to \textcolor{teal}{d_{2,0}}}   & \targ                        & \ctrl{1}                      & \qw      & \meter \\
        \lstick{d_{3} \to \textcolor{violet}{d_{3,0}}} & \qw                          & \targ                         & \ctrl{1} & \meter \\
        \lstick{d_{4} \to \textcolor{violet}{d_{3,1}}} & \qw                          & \targ                         & \targ    & \meter \\
        \lstick{d_{5} \to \textcolor{teal}{d_{2,1}}}   & \targ                        & \ctrl{-1}                     & \qw      & \meter \\
        \lstick{d_{6} \to \textcolor{cyan}{d_{1,1}}}   & \ctrl{-1}                    & \qw                           & \qw      & \meter \\
        }   
        \label{fig:cx_half}
    }\\
    \subfloat[]{
        \centering
        \Qcircuit @C=2em @R=1em {
        \lstick{\textcolor{cyan}{d_{1,0}}}   & \ctrl{2} & \qw \barrier[-1.5em]{5} & \qw      & \qw  \barrier[-1.5em]{5} & \qw      & \meter \\ 
        \lstick{\textcolor{cyan}{d_{1,1}}}   & \qw      & \ctrl{2}                & \qw      & \qw                      & \qw      & \meter \\
        \lstick{\textcolor{teal}{d_{2,0}}}   & \targ    & \qw                     & \ctrl{2} & \qw                      & \qw      & \meter \\
        \lstick{\textcolor{teal}{d_{2,1}}}   & \qw      & \targ                   & \qw      & \ctrl{2}                 & \qw      & \meter \\
        \lstick{\textcolor{violet}{d_{3,0}}} & \qw      & \qw                     & \targ    & \qw                      & \ctrl{1} & \meter \\
        \lstick{\textcolor{violet}{d_{3,1}}} & \qw      & \qw                     & \qw      & \targ                    & \targ    & \meter
        \gategroup{1}{1}{2}{1}{.5em}{\{}
        \gategroup{3}{1}{4}{1}{.5em}{\{}
        \gategroup{5}{1}{6}{1}{.5em}{\{}
        \\
        }
        \label{fig:cdcx}
    }
    \caption{An ``outside-in'' implementation of a CX ladder consisting of 5 CXs. (a) An implementation on a $6\times1$ line (or $3\times2$ grid) of qubits with nearest-neighbor connectivity. (b) A rearrangement of (a) via row-major ordering.  The CX ladder places the sum (modulo 2) of every individual qubit's state on the middle/last qubit for sub-figures (a)/(b) respectively. The dashed barriers delineate segments of the circuit that can be executed in parallel and the brackets denote qubits within the same row.}
\end{figure}

Figure \ref{fig:cdcx} shows the same circuit as Figure \ref{fig:cx_half} but with the double-indexed qubit labels rearranged to place the qubits within the same row next to one another.
One should think of this arrangement as two initial CX ladders being executed on the two columns of qubits in parallel followed by a single CX that adds their respective partial summations. 
Furthermore, and most importantly, this arrangement allows for a convenient logical encoding in which each row of qubits are designated to be the two logical qubits within a [[4, 2, 2]] patch. With this encoding, each CX simply acts transversally between patches (except for the last one, which acts within the final patch).

\subsection{To Constant Depth}

In this experiment, we run circuits that allow CX ladders to be executed in constant depth. Generally there exists a space-time tradeoff for quantum circuits that enables one to employ entanglement, mid-circuit measurement, and feed-forward operations to reduce circuit depth at the expense of requiring additional qubits \cite{baumer2024efficient, hashim2024efficient, Baumer2025}. 
This exchange can be quite favorable since decreasing the depth of a quantum circuit decreases the noise it suffers due to depolarization \cite{temme2017error}, and is particularly well-suited to neutral atom architectures that can support thousands of qubits within a single array \cite{saffman2019quantum}. 

To illustrate the reduction of a CX ladder to constant depth, we start with a small example: the 6-qubit ladder shown in Figure \ref{fig:cdcx}. We will reduce its depth by parallelizing its first two sections (where sections are delineated by dashed vertical lines), ultimately resulting in the circuit shown in Figure \ref{fig:cdcx_unencoded}.
\begin{figure}[t]
\centering
\[
\Qcircuit @C=1em @R=1em {
\lstick{} & \qw      & \qw      & \qw \barrier[0em]{7} & \qw & \ctrl{2} & \qw \barrier[0em]{7} & \qw & \qw      & \meter & \cw                   & \cw                       & \cw \\     
\lstick{} & \qw      & \qw      & \qw                  & \qw & \qw      & \ctrl{2}             & \qw & \qw      & \meter & \cw                   & \cw                       & \cw \\
\lstick{} & \gate{H} & \ctrl{2} & \qw                  & \qw & \targ    & \qw                  & \qw & \qw      & \meter & \cw                   & \cctrl{2}                       \\     
\lstick{} & \gate{H} & \qw      & \ctrl{2}             & \qw & \qw      & \targ                & \qw & \qw      & \meter & \cctrl{2}                                               \\
\lstick{} & \qw      & \targ    & \qw                  & \qw & \ctrl{2} & \qw                  & \qw & \qw      & \meter & \cw                   & \push{\oplus} \cw \cwx[2] & \cw \\
\lstick{} & \qw      & \qw      & \targ                & \qw & \qw      & \ctrl{2}             & \qw & \qw      & \meter & \push{\oplus} \cw \cwx[2] & \cw                   & \cw \\
\lstick{} & \qw      & \qw      & \qw                  & \qw & \targ    & \qw                  & \qw & \ctrl{1} & \meter & \cw                   & \push{\oplus} \cw \cwx[1] & \cw \\
\lstick{} & \qw      & \qw      & \qw                  & \qw & \qw      & \targ                & \qw & \targ    & \meter & \push{\oplus} \cw     & \push{\oplus} \cw         & \cw
\inputgroupv{1}{2}{0.8em}{0.8em}{\ket{00}_L\;\;}
\inputgroupv{3}{4}{0.8em}{0.8em}{\ket{00}_L\;\;}
\inputgroupv{5}{6}{0.8em}{0.8em}{\ket{00}_L\;\;}
\inputgroupv{7}{8}{0.8em}{0.8em}{\ket{00}_L\;\;}
\\
}
\]
\caption{Logical circuit for a 5 CX, constant-depth CX ladder on a $4\times2$ grid of qubits. 
The brackets surround qubits on the same row. 
The dashed lines separate the circuit into three parts: 
the first creates a Bell pair between the ancilla row and the second data row, 
the second consists of the CXs from the first two sections of the original circuit (Figure \ref{fig:cdcx}) being executed in parallel, 
and the third consists of the final CX from the original circuit, measurement, and classical correction.
}
\label{fig:cdcx_unencoded}
\end{figure}
This can be achieved by introducing a row of two ancilla qubits between the first two rows of data qubits. This ancilla row is then pair-wise entangled with the second data row by preparing them in a GHZ state: $(\ket{0000}+\ket{1111})/\sqrt{2}$. Because of this entanglement, it is as if the targets of the first CX pair (of the second section of Figure \ref{fig:cdcx_unencoded}) were applied not on the ancilla row, but the second data row, before the targets of the second CX pair. However, this is only true if the ancilla row is measured to be $\ket{00}$. If it is not, one could imagine that it would have been had the appropriate set of $X$ gates been applied immediately before measurement. For this circuit, these ``correcting'' gates can be pushed down the circuit and past the measurement where they can be implemented as bit flips in post-processing. The four rows of logical qubits are initialized to the logical $\ket{00}_L$ state of the [[4, 2, 2]] code (Equation \ref{eq:codewords}). Each of these initializations requires a fifth physical qubit acting as a flag qubit to ensure the fault tolerance of the preparation. To achieve the ring connectivity required for this preparation, the entire column of flag-qubit atoms is moved, mid-circuit, from the last column of the grid to the first, as depicted in the upper panel of Figure \ref{fig:summary_of_results}c(ii).
Figure \ref{fig:cdcx_unencoded} depicts the circuit at the logical level, and it is decomposed to the physical level in the [[4, 2, 2]] code, and further to Sqale's native gateset, via the decomposition found in the Appendix of our previous work \cite{bedalov2024fault}.

This logic may be extended to arbitrarily long CX ladders as shown in Figure \ref{fig:cdcx_arbitrary} of Appendix \ref{sec:cdcx_arbitrary}. In particular, the logical circuit for the 12 logical qubit experiment is simply this figure without the ellipses. In general, the post-processing is as follows: the measured bit of a data qubit is flipped if the measured bitstring of the previous ancilla qubits (along the same column) has even parity. For the last data qubit, the string of all previous ancilla qubits is considered.

Here we can achieve this logical, constant-depth scaling due to the use of $[[4,2,2]]$ patches where each patch sits in a row of atoms, as shown in Figure~\ref{fig:summary_of_results}. Movement costs are constant, as transversal CXs require no movement, and flag qubits only move a fixed horizontal distance. 
Going forward, we may also be interested in performing such circuits on logical qubits of arbitrary distance $d$, where each code block may require a 2D array of atoms whose width and height scale with $d$. When performing transversal CXs or state preparation, movement time will therefore scale with $d$. While this may suggest the loss of a constant-depth advantage, as we will discuss in Section~\ref{sec:motivation_constant} the logical, nearest-neighbor circuit structure mitigates movement costs at scale.

\subsection{Results}

\begin{figure}
    \centering
    \includegraphics[width=1\linewidth]{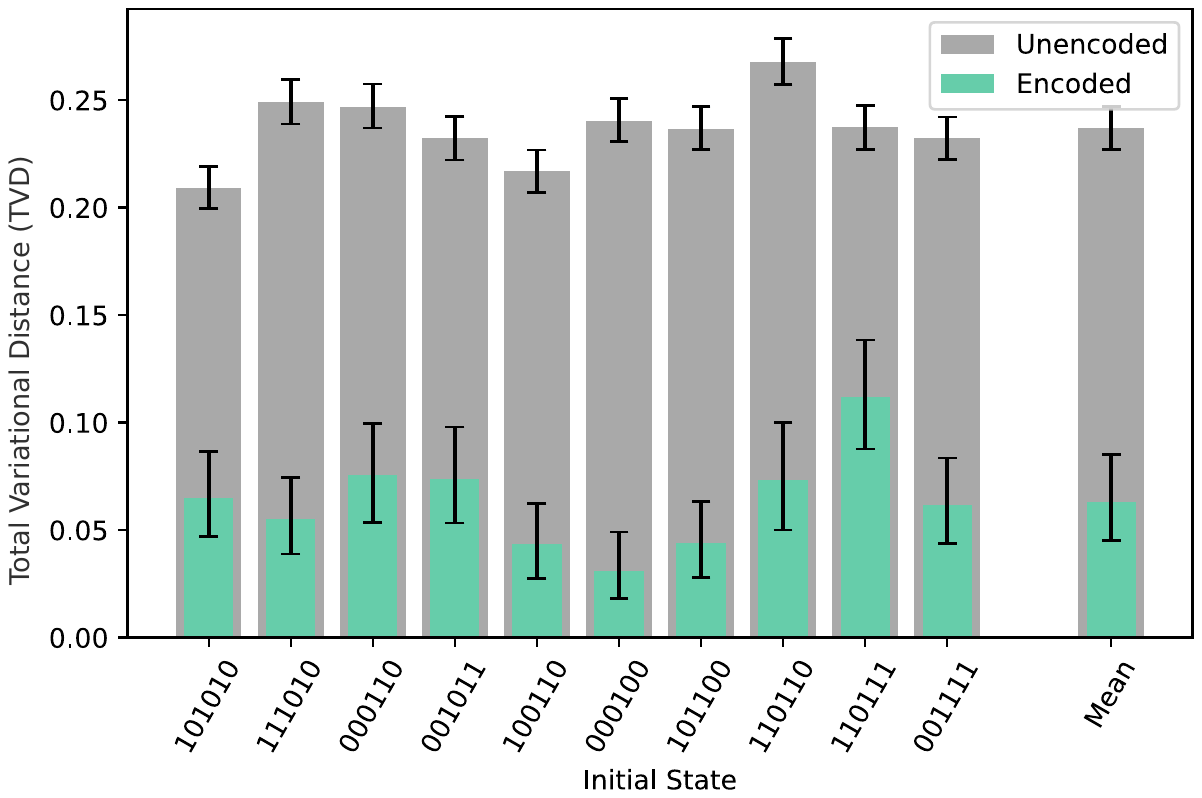}
    \caption{Error rates with 68\% confidence intervals for unencoded (gray, wide) vs encoded (green, narrow) constant-depth CX ladders with 8 logical qubits.}
    \label{fig:cdcx_8lq}
\end{figure}

\begin{figure}
    \centering
    \includegraphics[width=1\linewidth]{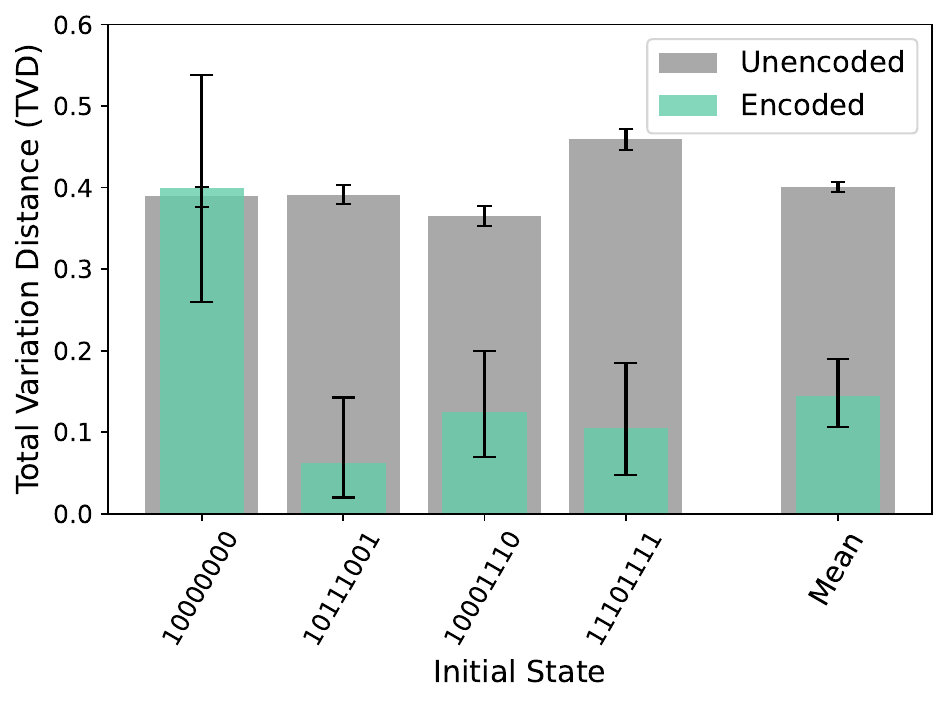}
    \caption{Error rates with 68\% confidence intervals for unencoded (gray, wide) vs encoded (green, narrow) constant-depth CX ladders with 12 logical qubits.}
    \label{fig:cdcx_12lq}
\end{figure}

We collected experimental results for both unencoded and encoded constant-depth CX ladders with both 8 and 12 logical qubits. 
The measurements of these circuits are post-processed via loss detection, decoding (for the encoded circuits), and Pauli correction (as described in Section \ref{subsection:post-processing}).
The unencoded and encoded 8 logical qubit circuits had a post-processing survival rate of 0.804 and 0.066, respectively. 
Meanwhile, the unencoded and encoded 12 logical qubit circuits had a post-processing survival rate of 0.768 and 0.009, respectively. 
Here we report success rates: the fraction of accepted measurements equal to the single, expected bitstring for each ladder.
The results for the 8 logical qubit experiment are given in Figure \ref{fig:cdcx_8lq}.
For this experiment, we collected data over 10 random initial product states. Every encoded circuit achieved a statistically significant decrease in error rate compared to its unencoded counterpart, with an average encoded error rate nearly 4 times lower than the unencoded error rate.
The results for the 12 logical qubit experiment are given in Figure \ref{fig:cdcx_12lq}.
For this experiment, we collected data over 4 random initial product states and found average encoded error rate more than 2 times lower than the unencoded error rate.

\section{Many-Hypercube Code}\label{sec:hypercube}

We can recursively concatenate the $[[4,2,2]]$ code introduced in Section~\ref{subsection:logical_encoding} with itself to obtain a family of high-rate and high threshold codes known as many-hypercube codes~\cite{goto2024high}.

In this section, we demonstrate the initialization of a level $L=2$ many-hypercube code derived from concatenating the $[[4,2,2]]$ code with itself once. 

We first write the stabilizers for the $[[4,2,2]]$ code as
\begin{align*}
    S_X^{(1)}= \begin{bmatrix}
        X & X & X & X
    \end{bmatrix},
    \quad
    S_Z^{(1)} = \begin{bmatrix}
        Z & Z & Z & Z
    \end{bmatrix},
\end{align*} 
and in a similar way write the logical operators as
\begin{align*}
    &\overline{X}_1^{(1)}=\begin{bmatrix}
        \id & X & X & \id
    \end{bmatrix},
    &&
    \overline{Z}_1^{(1)}=\begin{bmatrix}
        Z & \id & Z & \id
    \end{bmatrix}, \\
    &\overline{X}_2^{(1)}=\begin{bmatrix}
        \id & X & \id & X
    \end{bmatrix},
    &&
    \overline{Z}_2^{(1)}=\begin{bmatrix}
        Z & \id & \id & Z
    \end{bmatrix}.
\end{align*}
We can then write the six $X$ and $Z$ stabilizers for the $[[16,4,4]]$, $L=2$ many-hypercube code as:
\begin{align*}
    &S_X^{(2)} = \begin{bmatrix}
        S^{(1)}_X & \id & \id & \id \\
        \id & S^{(1)}_X & \id & \id \\
        \id & \id & S^{(1)}_X & \id \\
        \id & \id & \id & S^{(1)}_X \\
        \overline{X}_1^{(1)} & \overline{X}_1^{(1)} & \overline{X}_1^{(1)} & \overline{X}_1^{(1)}\\
        \overline{X}_2^{(1)} & \overline{X}_2^{(1)} & \overline{X}_2^{(1)} & \overline{X}_2^{(1)}
    \end{bmatrix},
    \\
    &S_Z^{(2)} = \begin{bmatrix}
        S^{(1)}_Z & \id & \id & \id \\
        \id & S^{(1)}_Z & \id & \id \\
        \id & \id & S^{(1)}_Z & \id \\
        \id & \id & \id & S^{(1)}_Z \\
        \overline{Z}_1^{(1)} & \overline{Z}_1^{(1)} & \overline{Z}_1^{(1)} & \overline{Z}_1^{(1)}\\
        \overline{Z}_2^{(1)} & \overline{Z}_2^{(1)} & \overline{Z}_2^{(1)} & \overline{Z}_2^{(1)}
    \end{bmatrix}, 
\end{align*}
and the logical operators as:
\begin{align}
    &\overline{X}_1^{(2)} = \begin{bmatrix}
        \id & \overline{X}^{(1)}_1 & \overline{X}^{(1)}_1 & \id
    \end{bmatrix},
    &&\overline{Z}_1^{(2)} = \begin{bmatrix}
        \overline{Z}^{(1)}_1 & \id & \overline{Z}^{(1)}_1 & \id
    \end{bmatrix},\nonumber\\
    &\overline{X}_2^{(2)} = \begin{bmatrix}
        \id & \overline{X}^{(1)}_1 & \id & \overline{X}^{(1)}_1
    \end{bmatrix},
    &&\overline{Z}_2^{(2)} = \begin{bmatrix}
        \overline{Z}^{(1)}_1 & \id & \id & \overline{Z}^{(1)}_1
    \end{bmatrix},\nonumber\\
    &\overline{X}_3^{(2)} = \begin{bmatrix}
        \id & \overline{X}^{(1)}_2 & \overline{X}^{(1)}_2 & \id
    \end{bmatrix},
    &&\overline{Z}_3^{(2)} = \begin{bmatrix}
        \overline{Z}^{(1)}_2 & \id & \overline{Z}^{(1)}_2 & \id
    \end{bmatrix},\nonumber\\
    &\overline{X}_4^{(2)} = \begin{bmatrix}
        \id & \overline{X}^{(1)}_2 & \id & \overline{X}^{(1)}_2
    \end{bmatrix},
    &&\overline{Z}_4^{(2)} = \begin{bmatrix}
        \overline{Z}^{(1)}_2 & \id & \id & \overline{Z}^{(1)}_2
    \end{bmatrix}.
    \label{eq:mhc-stabilizers}
\end{align}

We verify this indeed corresponds to a code distance $d=4$ with the qLDPC library from Infleqtion and open-source collaborators~\cite{perlin2023qldpc}.

\subsection{Logical Circuits}

To concatenate the [[4, 2, 2]] code with itself, we prepare five [[4, 2, 2]] logical patches in the $\ket{00}$ state, where each patch is a row of four data qubits and an additional flag qubit. We move the last column of flag qubits to complete the fault-tolerant state preparation routine. We then perform the same state preparation circuit at the logical level: The first four patches become eight logical data qubits and the fifth patch provides two logical flag qubits. This requires moving the logical flag qubit row, as depicted in Fig. \ref{fig:summary_of_results}c(i). The result is a $\ket{0000}$ state in the [[16, 4, 4]] code. To prepare specific states, we apply transversal $X$ gates to the relevant qubits. Descriptions of the relevant fault-tolerant state preparation and transversal operations can be found in \cite{bedalov2024fault}.

\begin{figure}
    \centering
    \[
        \Qcircuit @R=1em @C=0.75em {
     \\
     &\lstick{}& \qw&                \qw&         \qw    &         \qw    &\targ    \qw    &         \qw    &         \qw    &         \qw    &\control \qw    &         \qw    &\meter\\
     &\lstick{}& \qw&                \qw&         \qw    &         \qw    &         \qw\qwx&\targ    \qw    &         \qw    &         \qw    &         \qw\qwx&\control \qw    &       \meter\\
     &\lstick{}& \qw&\gate{H} \qw&\control \qw    &         \qw    &\control \qw\qwx&         \qw\qwx&         \qw    &         \qw    &         \qw\qwx&         \qw\qwx&       \meter\\
     &\lstick{}& \qw&\gate{H} \qw&         \qw\qwx&\control \qw    &         \qw    &\control \qw\qwx&         \qw    &         \qw    &         \qw\qwx&         \qw\qwx&       \meter\\
     &\lstick{}& \qw&                \qw&\targ    \qw\qwx&         \qw\qwx&\control \qw    &         \qw    &         \qw    &         \qw    &         \qw\qwx&         \qw\qwx&       \meter\\
     &\lstick{}& \qw&                \qw&         \qw    &\targ    \qw\qwx&         \qw\qwx&\control \qw    &         \qw    &         \qw    &         \qw\qwx&         \qw\qwx&       \meter\\
     &\lstick{}& \qw&                \qw&         \qw    &         \qw    &\targ    \qw\qwx&         \qw\qwx&\control \qw    &         \qw    &         \qw\qwx&         \qw\qwx&       \meter\\
     &\lstick{}& \qw&                \qw&         \qw    &         \qw    &         \qw    &\targ    \qw\qwx&         \qw\qwx&\control \qw    &         \qw\qwx&         \qw\qwx&       \meter\\
     &\lstick{}& \qw&                \qw&         \qw    &         \qw    &         \qw    &         \qw    &\targ    \qw\qwx&         \qw\qwx&\targ    \qw\qwx&         \qw\qwx&       \meter\\
     &\lstick{}& \qw&                \qw&         \qw    &         \qw    &         \qw    &         \qw    &         \qw    &\targ    \qw\qwx&         \qw    &\targ    \qw\qwx&       \meter
     \inputgroupv{2}{3}{0.8em}{0.8em}{\ket{00}_L\;\;}
     \inputgroupv{4}{5}{0.8em}{0.8em}{\ket{00}_L\;\;}
     \inputgroupv{6}{7}{0.8em}{0.8em}{\ket{00}_L\;\;}
     \inputgroupv{8}{9}{0.8em}{0.8em}{\ket{00}_L\;\;}
     \inputgroupv{10}{11}{0.8em}{0.8em}{\ket{00}_L\;\;}
     \\
    }
    \]
    \caption{$L=2$ many-hypercube state preparation circuit. The five pairs of qubits are initialized to the logical $\ket{00}_L$ state of the [[4, 2, 2]] code (Equation \ref{eq:codewords}).}
    \label{fig:placeholder}
\end{figure}

\subsection{Results}

\begin{figure}[H]
    \centering
    \includegraphics[width=.9\linewidth]{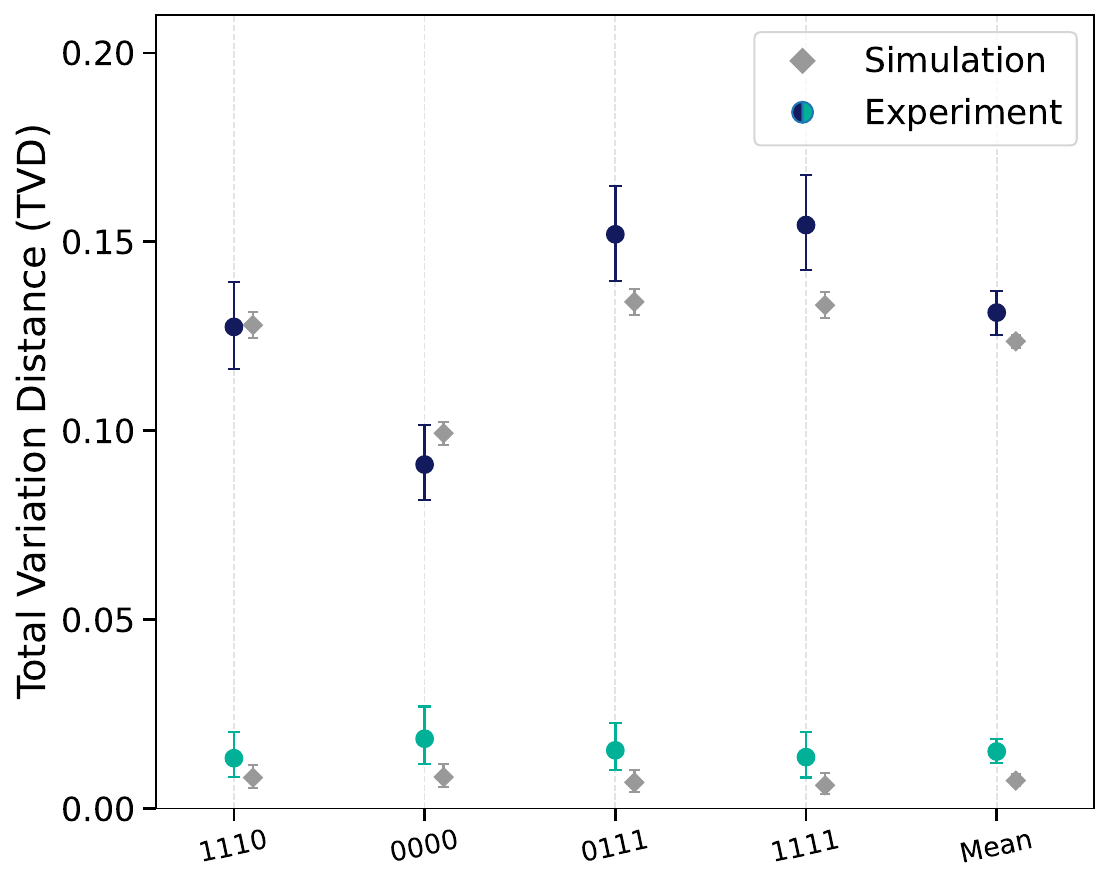}
    \caption{Experimental demonstration of the many-hypercube state preparation for various initial states. The plot compares error rate (with error bars for a 68\% confidence interval) for unencoded (dark blue) vs. encoded (green) state preparation for an assortment of bitstrings. Simulated values (gray diamonds) are shown next to each measured result (generated using Sqalesim with error factor 1).}
    \label{fig:mhc_results}
\end{figure}

We collected experimental results for unencoded and encoded ([[16, 4, 4]]) state preparation of the bitstrings 1110, 0000, 0111, and 1111. As usual, the results from the unencoded circuits were post-selected on atom loss. In this case, post-processing of the encoded circuits included both post-selection and error correction.
Stabilizer values are reconstructed from data qubit measurements at the end of the circuit. As a result, only $Z$ stabilizers are used for error rejection in decoding.
We post-select on the zero state of the (physical) flag qubits from the five [[4, 2, 2]] codes prepared in the first level of encoding, and the $\ket{00}_L$ state of the logical ancilla at the next level. For a distance-4 QEC code, we can correct a single error on any data qubit, and further detect (but not always correct) any pair of errors; at least three single-qubit errors are required for the decoder to miscorrect into an erroneous logical state. With similar confidence, we can correct pairs of atom losses (as long as we don't attempt to correct any other errors in the same sample). We can simultaneously correct up to one single atom loss when decoding the logical ancilla while preserving our overall confidence (because we are post selecting on a single logical outcome, at least two more single-qubit errors would be required to produce a false negative).
Across all bitstrings, the encoded circuits have a lower error rate than the unencoded ones. We measure an average encoded error rate of 0.0151, a significant improvement relative to the 0.1312 unencoded error rate. 
We also compute the average error rate by collecting all of the correct and incorrect counts across the four bitstrings and comparing them to the ideal case (where all the counts correspond to the correct bitstring). The mean postselection rate for the encoded circuits is 8.65\%, while the unencoded circuits have no atom loss events. The full result can be seen in Figure \ref{fig:mhc_results}.

\section{Towards a Fault-Tolerant Architecture}
\label{sec:outlook}
\label{sec:motivation_constant}
While the experimental demonstrations in this work realized transversal CXs on $[[4,2,2]]$ codes, we can implement multi-qubit logical gates using transversal CXs and atom movement for CSS codes of arbitrary distance $d$.
Additionally for platforms such as neutral atoms where space-time tradeoffs favor reduced time costs, an appealing target is a constant-depth logical architecture~\cite{zheng2020constant}.
Motivated by these ideas, we now describe how the features of this work fit into a logical architecture for constant-depth circuits at scale. 

\subsection{Key System Features}
We first point out that two hardware upgrades to Sqale: parallel gates and mid-circuit measurements, will be needed to realize a constant-depth architecture. 
Notably these features have been successfully realized in other neutral atom architectures~\cite{bluvstein2025logical}.
 
We also point out that two existing mechanisms demonstrated in this work: syndrome decoding and atom movement, have time costs that can scale as system size increases.
Unmitigated, this can erase the benefit of a constant-depth logical architecture.
Here we discuss how our focus on logical constant-depth circuits 
is beneficial and reduces the scaling costs of syndrome decoding and atom movement.



\subsubsection{Syndrome Decoding}
Popular code deformation techniques for performing logic~\cite{swaroop2024universal, cohen2022low, horsman2012surface} require a number of syndrome measurement rounds scaling as $O(d)$ to handle measurement errors properly.
In contrast, transversal gates can require $O(1)$ syndrome measurement rounds so long as proper decoding techniques are used~\cite{zhou2024algorithmic, serra2025decoding, cain2024decoding}.
These decoding techniques, however, require a decoding volume scaling with the number of logical qubits included in the \textit{observing region} for an $O(d)$ round decoding window.
For a worst-case circuit, such as a depth-$O(d)$ tree of CXs, the observing region can contain $2^{O(d)}$ logical qubits.
In such a case, stalls from decoding complexity can erase a constant-depth advantage.
In contrast, our approach only uses constant, near-neighbor logical movement operations for multi-qubit logical operations.
The number of logical qubits that can be in the observing region after $O(d)$ rounds is therefore at most $O(d^2)$, allowing us to avoid a worst-case exponential increase in decoding complexity.


\subsubsection{Movement Operations}
Although the use of transversal CXs reduces the syndrome measurement overhead to $O(1)$, the physical implementation of a transversal CX relies on atom movement that scales with both the code block size and the interblock distance. 
In the simplest case we can assume an architecture created by tiles of code blocks on a 2D grid, such as those studied for surface code computation~\cite{litinski2019gameofsurfacecodes, sunami2025}, where each code block has physical size $O(d^2)$.
For arbitrary circuits on $N$ logical qubits, the worst-case movement cost would be $O(\sqrt{N}d)$. 
However, for the constant-depth circuits we consider all logical operations are between near neighbor qubits which corresponds to an $O(1)$ interblock distance.
The worst-case movement costs for a transversal CX in our approach is therefore $O(d)$.
Furthermore, by using the minimum-jerk movement trajectory demonstrated in this work (Appendix~\ref{sec:app_rearrange}), the wall-clock time of movement operations for a transversal CX scales as $O(d^{\frac{1}{3}})$.





\section{Conclusion} \label{sec:conclusion}
This work presents a neutral atom quantum computing architecture that unites qubit motion with individual optical addressing to run logical quantum circuits. This architecture reduces the overhead of atom motion that might otherwise hamper the scaling of utility-scale quantum computers~\cite{sunami2025}.
The experimental demonstrations exhibit improved logical-over-physical performance for the first demonstrations of logical pre-compiled Shor circuits and logical constant-depth circuits.

Two complementary ingredients underpin these results. First,  manual circuit compilation to achieve constant depth and optimized qubit placement. Second, in‑place operations enable fast entanglement without requiring movement before every gate, mitigating a dominant contributor to wall‑clock runtime and error accumulation.
Together, these features enable a neutral atom architecture that seeks to minimize circuit depth and runtime at the expense of increased qubit count:. This is a trade‑off well matched to neutral atom systems, which can increase qubit number more readily than they can accelerate native gate times.

Looking ahead, the current single‑species implementation described in this work would still require motion for mid‑circuit measurements; extending to dual‑species arrays or hiding beams would enable in‑place readout and further reduce both runtime and error \cite{fang2025interleaved, saffman2025quantum, sunami2025}.

Continued development in several key areas will be central to exploiting constant‑depth paradigms at larger scale: optical addressing for parallel gates, scaled tuning systems for individually addressed entangling operations, improvement in SPAM and gate fidelities, and continuous atom reloading \cite{chiu2025continuousoperationofacoherent3, muniz2025repeatedancillareuselogical, li2025fastcontinuouscoherentatom}. Additionally, scaling the physical qubit count to tens and hundreds of thousands is essential and achievable, as demonstrated by recent progress towards a 6,100-qubit array \cite{manetsch2025tweezerarray6100highly, chiu2025continuousoperationofacoherent3}.

By consolidating recent hardware and compilation advances, this architecture demonstration lays the groundwork for further increases in scale towards large, fault-tolerant quantum computers.
\section{Acknowledgements}\label{sec:acknowledgements}

We thank Cord Mazzetti, Sayam Sethi, and Jonathan Mark Baker for their discussions regarding early fault-tolerant implementations of Shor’s algorithm.

This material is based upon work supported by the U.S. Department of Energy, Office of Science, Office of Advanced Scientific Computing Research, under Award Numbers DE-SC0021526 and DE-SC0025493. Work on this manuscript is supported by Wellcome Leap as part of the ‘Quantum Biomarker Algorithms for Multimodal Cancer Data’ research project within the Quantum for Bio (Q4Bio) Program. 
This work is also funded in part by the STAQ project under award NSF Phy-232580; in part by the US Department of Energy Office of Advanced Scientific Computing Research, Accelerated 
Research for Quantum Computing Program; and in part by the NSF Quantum Leap Challenge Institute for Hybrid Quantum Architectures and Networks (NSF Award 2016136).  This work was completed in part with resources provided by the University of Chicago’s Research Computing Center.

Disclaimer: This report was prepared as an account of work sponsored by an agency of the United States Government.  Neither the United States Government nor any agency thereof, nor any of their employees, makes any warranty, express or implied, or assumes any legal liability or responsibility for the accuracy, completeness, or usefulness of any information, apparatus, product, or process disclosed, or represents that its use would not infringe privately owned rights.  Reference herein to any specific commercial product, process, or service by trade name, trademark, manufacturer, or otherwise does not necessarily constitute or imply its endorsement, recommendation, or favoring by the United States Government or any agency thereof.  The views and opinions of authors expressed herein do not necessarily state or reflect those of the United States Government or any agency thereof.
\section{Interests}\label{sec:interests}
The authors declare no competing non-financial interests but the following financial interest: Rich Rines, Benjamin Hall, Joshua Viszlai, Daniel C. Cole, David Mason, Cameron Barker, Matt J. Bedalov, Matt Blakely, Tobias Bothwell, Caitlin Carnahan, Frederic T. Chong, Samuel Y. Eubanks, Brian Fields, Matthew Gillette, Palash Goiporia, Pranav Gokhale, Garrett T. Hickman, Marin Iliev, Eric B. Jones, Ryan A. Jones, Kevin W. Kuper, Stephanie Lee, Martin T. Lichtman, Kevin Loeffler, Nate Mackintosh, Farhad Majdeteimouri, Peter T. Mitchell, Thomas W. Noel, Ely Novakoski, Victory Omole, David Owusu-Antwi, Alexander G. Radnaev, Anthony Reiter, Mark Saffman, Bharath Thotakura, Teague Tomesh, and Ilya Vinogradov are employees and/or shareholders of Infleqtion, Inc. Frederic T. Chong is an advisor to Quantum Circuits, Inc.
\section{Data Availability}\label{sec:data}
Data supporting the findings of this study are available on Zenodo at https://zenodo.org/records/17137995
\clearpage
\section{Appendix}
\label{sec:app1}

\subsection{Hardware System}\label{sec:app_hardware}
These results were obtained on Infleqtion's Sqale QPU. This system was previously described in Ref. \cite{radnaev2024universal} and was used for demonstrations described in Ref. \cite{bedalov2024fault}. The system has been upgraded, with changes including (a) replacement of crossed acousto-optic deflectors (AODs) by a spatial light modulator for generation of the static optical tweezer array and (b) addition of a 1040 nm optical tweezer system for mid-circuit rearrangement. The latter is directed by one of the crossed AOD pairs that is used for Rydberg addressing, with the larger (smaller in the trap focal plane) rearrangement beam and the smaller (larger in the focal plane) Rydberg beam time-multiplexed into the crossed AODs by upstream acousto-optic modulators.

\subsubsection{Gate Fidelities}

Single-qubit gates across the qubits used in this work were characterized as described in Ref. \cite{radnaev2024universal}, and measured fidelities were 99.80 (5) \% and 99.96 (1) \% for local (individually laser-addressed) and global (microwave-driven) single-qubit gates, respectively. Quoted uncertainties are standard deviations of the entire set of measured fidelities, aggregated across multiple runs of characterization of all qubits. 

Detailed randomized benchmarking of all entangling gates used in this work was not performed contemporaneously with data collection. Past characterization of our calibration procedure has shown that entangling gate fidelities across the qubits used in this work are typically between 98 and 99.5 \% after postselecting away atom loss. By approximating fidelity from single-circuit characterizations of entangling gates (as used for calibration as described in \cite{radnaev2024universal}), we estimate a postselected fidelity median of 98.7 \%, with an ensemble standard deviation of 0.3 \%.

\subsubsection{Mid-circuit Rearrangement}\label{sec:app_rearrange}

Multitone RF pulses applied to the AOD steering system allow several atoms to be transported in parallel. To avoid undesired interaction between the static and mid-circuit rearrangement (MCR) tweezer systems during atom transport, trajectories follow edges on the dual-lattice defined relative to the static tweezer array, i.e. the channels in-between the stationary tweezer beams. Movements proceed in five phases: first a simultaneous decrease in static trap depth and increase in MCR depth at the origin site, second a small diagonal displacement of the mobile atom onto a nearby dual-lattice site, third the long-distance transport of the atom to a dual-lattice site neighboring the destination site, and finally a reversal of the second then first steps to transfer the atom to the static destination tweezer. These displacement trajectories are exemplified in Fig. \ref{fig:summary_of_results}a. Each displacement follows a minimum-jerk trajectory, which mitigates excess heating of atoms in the regime of trap-depths and rastering speeds we target\cite{carruthers1965coherent, Chinnarasu2025}. Typical total movement durations are 7 ms and a single spin-echo microwave pulse is applied mid-movement.

We characterized MCR operations separately from the larger circuits described in this work. Averaging across all movements required for these demonstrations, individual atoms were successfully transferred to the static destination tweezer 98.0(8)\% of the time, and failures for atoms moved in parallel appear to be uncorrelated. We characterize the coherence of motion operations by inserting the motion gates into a Ramsey sequence. Correcting for state preparation and measurement error and virtually accounting for the frame flip induced by the spin-echo pulse, we find that movement operations apply the identity gate with 97.3(20) \% fidelity on moved atoms and with 98.3(19) \% fidelity on spectators. Preliminary results indicate that coherence improves significantly when the single spin-echo pulse is replaced by a more sophisticated dynamical decoupling sequence. The temperature of transported atoms increased from 5.1(2) $\mu$K prior to motion to 9.1(5) $\mu$K after motion; no appreciable heating of spectator atoms was observed. 

\subsection{System error model}

Table \ref{tab:sim_noise_model} summarizes the fidelities and error models for each operation included in the circuit-level simulator used to predict experimental results.

\begin{table} [h!]
\raggedright
\caption{Simulation Noise Model}
    \begin{tabular}{|c|m{0.85\columnwidth}|} \hline
         Operation& Error Mechanisms\\ \hline \hline
         
         State Prep & 
         \begin{minipage}[c][3.3cm]{0.85 \columnwidth} \raggedright
            \begin{itemize}
                \item Fidelity 97.9 \%
                 \item Initial state density matrix with $p=0.007$ and $q=0.014$: \\
                 \vspace{-0.25cm}
                 $$
                 \begin{pmatrix}
                 \frac{p}{2} & 0 & 0 & 0 & 0 \\
                 0 & 1-p-q & 0 & 0 & 0 \\
                 0 & 0 & \frac{p}{2} & 0 & 0 \\
                 0 & 0 & 0 & q & 0 \\
                 0 & 0 & 0 & 0 & 0 \\
                 \end{pmatrix}
                 $$
                 \end{itemize}
         \end{minipage}
         \\ \hline
         
         CZ &
         \begin{minipage}[c][4.1cm][c]{0.85 \columnwidth}
             \begin{itemize}
                \item Postselected fidelity 98.7 \%
                 \item Single qubit phase error with 0.73\% probability 
                 \item Spontaneous transitions with probabilities\\
                                  \vspace{-0.35cm}
                 $$
                 \begin{pmatrix}
                 17.4 & 185.0 & 4.9 & 165.4 & 0 \\
                 18.5 & 197.5 & 4.6 & 177.7 & 0 \\
                 31.1 & 420.3 & 45.8 & 1210.0 & 0 \\
                 42.1 & 590.2 & 52.9 & 1901.0 & 0 \\
                 0 & 5000 & 0 & 0 & 0 \\
                 \end{pmatrix} \times 10^{-6}
                 $$
             \end{itemize}
         \end{minipage}
         \\ \hline
         
         GR& 
         \begin{minipage}[c][1cm][c]{0.85 \columnwidth}
             \begin{itemize}
             \item Fidelity 99.96 \%
            \item Static overrotation of 0.0490 radians
             \end{itemize}
        \end{minipage}
         \\ \hline
         
         $\text{R}_\text{Z}$& 
         \begin{minipage}[c][3.5cm][c]{0.85 \columnwidth}
             \begin{itemize}
                 \item Fidelity 99.8 \%
                 \item Relative overrotation of $2.6\%$
                 \item Spontaneous transitions with probabilities\\
                 \vspace{-0.25cm}
                 $$
                 \begin{pmatrix}
                 0 & 207.2 & 0 & 0 & 0 \\
                 0 & 223.1 & 0 & 0 & 0 \\
                 0 & 364.3 & 0 & 0 & 0 \\
                 0 & 524.0 & 0 & 0 & 0 \\
                 0 & 0 & 0 & 0 & 0 \\
                 \end{pmatrix} \times 10^{-6}
                 $$
             \end{itemize}
        \end{minipage}
         \\ \hline
         
         Measure& 
         \begin{minipage}[c][6.3cm][c]{0.85 \columnwidth}
             \begin{itemize}
                \item Postselected fidelity 98.7 \%
                 \item Classification error:
                 \vspace{-0.23cm}
                 $$\text{Prob}(|i\rangle) = \text{Tr}(\mathcal{P}_i \rho)$$ where for $\epsilon_0 = 0.002$ and $\epsilon_1 = 0.023$
                 \vspace{-0.23cm}
                 $$ \mathcal{P}_{0} = 
                 \begin{pmatrix}
                 1-\epsilon_0 & 0 & 0 & 0 & 0 \\
                 0 & \epsilon_1 & 0 & 0 & 0 \\
                 0 & 0 & 1-\epsilon_0 & 0 & 0 \\
                 0 & 0 & 0 & \epsilon_1 & 0 \\
                 0 & 0 & 0 & 0 & 0 \\
                 \end{pmatrix} 
                 $$
                 $$
                \mathcal{P}_{1} = 
                 \begin{pmatrix}
                 \epsilon_0 & 0 & 0 & 0 & 0 \\
                 0 & 1-\epsilon_1 & 0 & 0 & 0 \\
                 0 & 0 & \epsilon_0 & 0 & 0 \\
                 0 & 0 & 0 & 1-\epsilon_1 & 0 \\
                 0 & 0 & 0 & 0 & 0 \\
                 \end{pmatrix} 
                 $$
             \end{itemize}
        \end{minipage}
         \\  \hline
         Movement& 
         \begin{minipage}[c][1.8cm][c]{0.85 \columnwidth}
             \begin{itemize}
                 \item Spectator phase error probability 2.6 \%; fidelity 98.3 \%
                 \item Movement phase error probability 4.1 \%; postselected fidelity 97.3 \%
             \end{itemize}
        \end{minipage}
         \\ \hline
    \end{tabular}

    *All matrices expressed in the $\{|0\rangle, |1\rangle, |0_\ell\rangle, |1_\ell\rangle, |L\rangle \}$ basis
    
    \label{tab:sim_noise_model}
\end{table}

\subsection{Error Factor Simulations}\label{sec:simulations}
\subsubsection{Shor}\label{sec:shor_noisy_simulations}
\begin{figure}
    \centering
    \includegraphics[width=\linewidth]{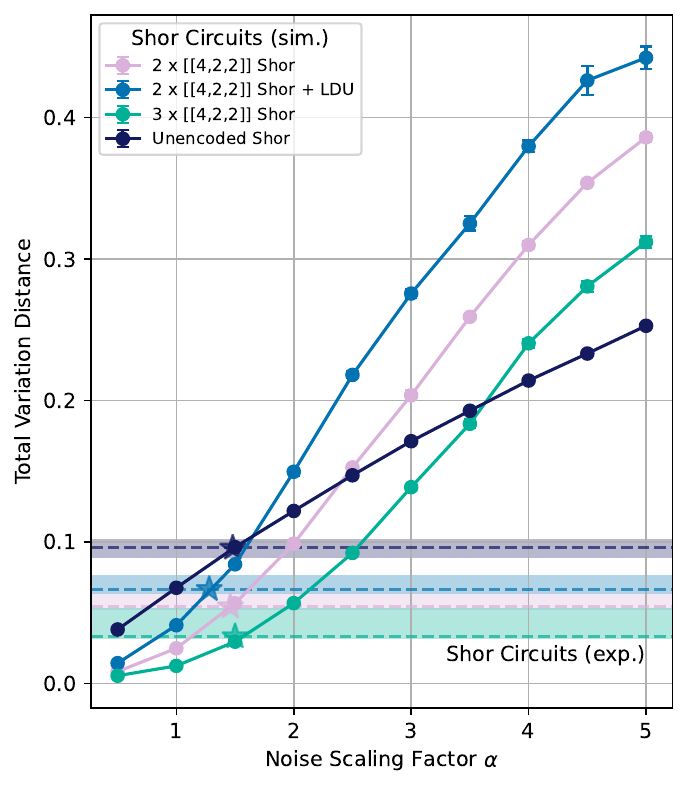}
    \caption{Noisy simulation results for all executed Shor circuits using Sqalesim. A subset of dominant errors are scaled by a factor $\alpha$ as a proxy for extrapolated noise scaling on hardware. The plot shows the simulated TVD computed against the ideal distribution. Each data point is the mean value across five repeated trials, with each trial using 6$\times 10^{4}$ shots for a given $\alpha$. Error bars shown are the standard error of the mean, and, where not visible, are smaller than the marker. For comparison, the mean TVDs found in experiment are shown using dashed horizontal lines, with the surrounding shaded area representing the confidence interval. Their intersections with the simulated curves are marked by stars that lie between $\alpha=1$ and $\alpha = 1.5$}
    \label{fig:shor_noise_sim_compare}
\end{figure}
We investigated the impact and sensitivity of hardware noise further for the circuits in \cref{sec:shor} by employing noise scaling analysis to discover their pseudo-thresholding behavior (see Fig. \ref{fig:shor_noise_sim_compare}). We identified the error sources in Sqalesim most likely to have higher hardware drift -- CZ single qubit phase error, $\text{R}_\text{Z}$ relative overrotation, movement and spectator phase error, and state preparation and measurement errors -- and scaled them by a multiplicative factor $\alpha$ (where the default settings correspond to $\alpha=1$). We expect the hardware to drift between an $\alpha$ value of 0.5 to 1.5 -- indeed, we see that the TVDs found in experiment intersect the simulated TVDs between $\alpha =1$ and $\alpha = 1.5$, suggesting that the hardware is well-characterized by the noise model and estimated noise parameters.

In taking 6$\times 10^{4}$ raw noisy shots to approximate the infinite shot limit, we observed that the $[[4,2,2]]$ encoded Shor circuits maintain a lower TVD compared to the unencoded Shor for all simulated values up to $\alpha\sim2.4$ and $\alpha\sim3.6$ for two- and three-row encodings, respectively. It is also interesting to see that there is a clear TVD separation between the two- and three-row Shor circuits themselves (as also observed experimentally in Fig. \ref{fig:shor_experimental_data}) for all $\alpha$ we simulated above $\sim$0.5. We attribute this general resilience to the strong noise robustness of the $[[4 2,2]]$ code for early fault-tolerant circuits \cite{bedalov2024fault}, with the longer sustained performance of the three-row encoding arising from more opportunities for error detection filtering.
This highlights that different logical encodings of the same underlying code can exhibit observable differences in the presence of realistic noise, and that a balance exists between a compact encoding and broadly distributed logical redundancy. At the same time, we also notice the potential limitations of the $[[4,2,2]]$ code with the simulations of the two-row Shor circuit equipped with LDU. Its pseudothreshold (the point at which unencoded and encoded performance cross) is $\alpha\sim1.7$, lower than its non-LDU counterpart. We explain this behavior as a result of applying numerous CZ gates for the LDU while CZ error accumulates (along with a proportional atom loss) with increasing $\alpha$. In particular, because the $[[4,2,2]]$ code has only distance 2, increasing $\alpha$ very quickly leads to logical errors, and whatever additional error detection capability is offered by the LDU gets overwhelmed. We note that this is particularly exemplified for the Shor circuits as the number of CZs added from the LDUs approaches the CZ depth of the base encoded circuit itself. Although the circuits with leakage detection perform better than the unencoded Shor circuits in the error regime $\alpha \leq 1$, these findings emphasize that informed decisions about the kinds of fault-tolerance gadgets to implement for a given circuit should be based on the dominant hardware error sources at play.

\subsubsection{Constant-depth CX ladder}

We also studied the simulated performance of constant-depth CX (CDCX) ladder circuits as we scale the identified noise parameters by  $\alpha = 0.1$ to $8$. Due to the number of shots required ($>5\times10^5$ for higher scaling factors), these simulations reflect results for a specific initial bitstring for each CDCX ladder size--initial simulations of multiple bitstrings saw little variation in results, suggesting that the analysis here should hold qualitatively across all initial bitstrings. 

\begin{figure}
    \centering
    a.\\
    \includegraphics[width=\linewidth, height=6cm]{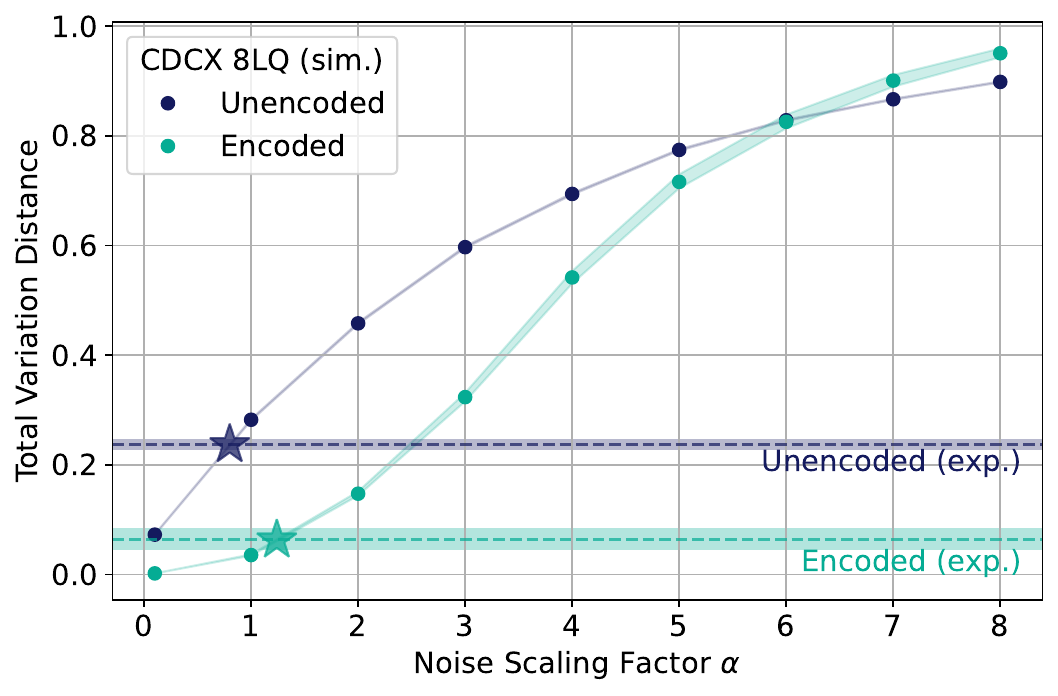} 
    \label{fig:cdcxsim_8lq_tvd}
    b.\\
    \includegraphics[width=\linewidth, height=6cm]{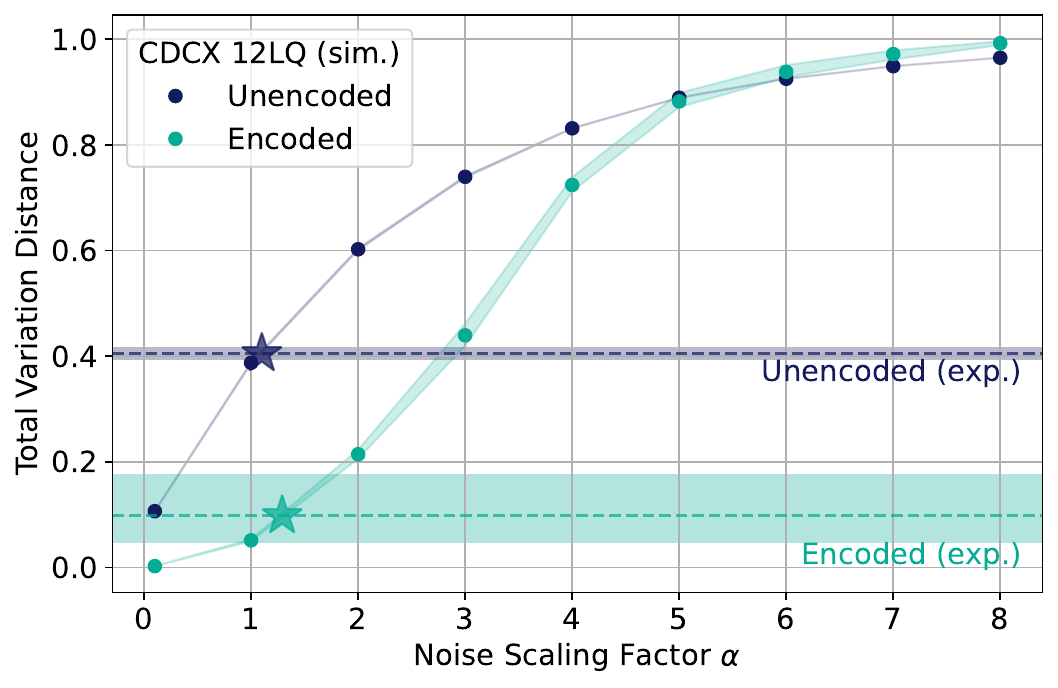}
    \label{fig:cdcxsim_12lq_tvd}
    \caption{Simulations of 8 LQ and 12 LQ CDCX circuits while scaling selected noise parameters. The horizontal lines show experimental results, and their intersections with the simulated TVDs are marked by stars.}
    \label{fig:cdcx_sim_tvd}
\end{figure}

Fig.~\ref{fig:cdcx_sim_tvd} shows the simulated TVDs of the encoded and unencoded circuits for 8 LQ and 12 LQ CDCX circuits. Up to a pseudothreshold of $\alpha\sim 6$ and $\alpha \sim 5$, respectively, the encoded circuit has a lower TVD than the unencoded circuit, after which the encoded circuit has a slightly higher TVD than the unencoded circuit.

For comparison, we also mark the average TVDs found for these circuits in experiment. We observe that these results intersect with the simulated curves between $\alpha \sim 0.8$ and $\alpha \sim 1.3$. The fact that experimental data matches simulation for $\alpha\sim0.8 - 1.1$ and $\alpha\sim1.2-1.3$ for unencoded and encoded results, respectively, may be explained by depth-dependent performance and/or temporal drift of gate fidelities.

\begin{figure}
    \centering
    \includegraphics[width=.95\linewidth]{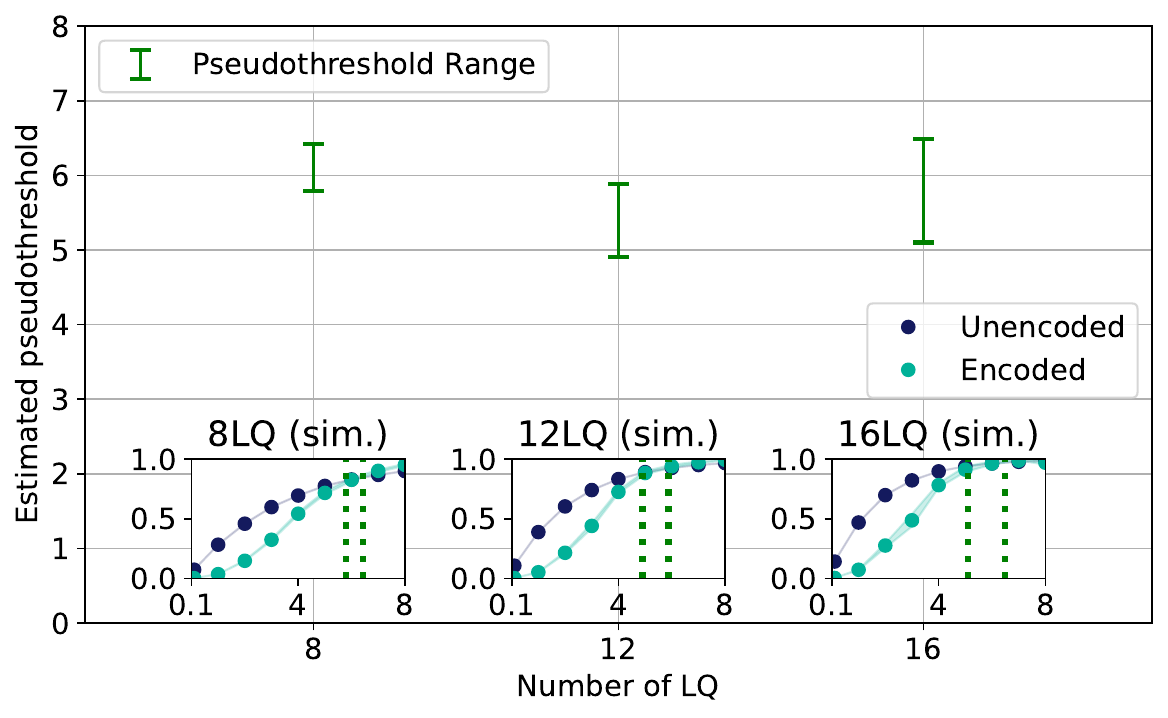}
    
    \caption{Scaling of estimated pseudothreshold (error factor at which encoded and unencoded circuit TVDs cross). The insets show the TVDs for the encoded and unencoded circuits as a subset of dominant error parameters are scaled, and vertical dotted lines indicating the estimated pseudothreshold range. }
    \label{fig:cdcxsim_scalinglq}
\end{figure}

Finally, we study performance as the number of logical qubits $N$ in the CDCX circuits is increased, characterizing performance using the estimated pseudothreshold (Fig.~\ref{fig:cdcxsim_scalinglq}). For each $N$, we find the range of noise scaling factors where the confidence intervals of the encoded and unencoded TVDs intersect and identify this as the pseudothreshold range. As $N$ increases, the depth of the CDCX ladder circuit increases, such that we might expect the need for lower error rates to see the benefits of encoding, ie. a decreased pseudothreshold. In simulation, we see a pseudothreshold range of around $5.8-6.4$ for $N=8$, $4.9 - 5.9$ for $N=12$, and $5.1-6.5$ for $N=16$.  
While the current device hardware is well below the pseudothreshold for the studied circuits, future applications involving more logical qubits may require moving towards error-correcting codes like the [[16, 4, 4]] many-hypercube code.

\subsection{Many-hypercube code simulation}
\label{subsection:mhc-sim}
Fig. \ref{fig:mhc_results} includes the results of noisy simulations of the [[16,4,4]] state preparation circuits performed using Sqalesim, compared against the experimental results. The encoded data is simulated using default parameters (Table \ref{tab:sim_noise_model}). The unencoded circuits, which use only four qubits, are simulated using qubit-specific characterization of state preparation and measurement error, which vary across the qubit array.  The simulations reproduce the experimental observation (Fig. \ref{fig:mhc_results}) of a pronounced separation in error rates between encoded and unencoded bitstrings.


\subsection{Arbitrary Logical CDCX Circuit}\label{sec:cdcx_arbitrary}
\begin{figure}[H]
    \centering
    \[
    \Qcircuit @C=1em @R=1em {
    \lstick{}                                 & \qw      & \qw                        & \qw \barrier[0em]{15}      & \qw & \ctrl{2} & \qw \barrier[0em]{15} & \qw & \qw      & \meter & \cw                        & \cw                        & \cw \\     
    \lstick{}                                 & \qw      & \qw                        & \qw                        & \qw & \qw      & \ctrl{2}              & \qw & \qw      & \meter & \cw                        & \cw                        & \cw \\
    \lstick{}                                 & \gate{H} & \ctrl{2}                   & \qw                        & \qw & \targ    & \qw                   & \qw & \qw      & \meter & \cw                        & \push{\oplus}\cw \cwx[2]   &     \\     
    \lstick{}                                 & \gate{H} & \qw                        & \ctrl{2}                   & \qw & \qw      & \targ                 & \qw & \qw      & \meter & \push{\oplus}\cw \cwx[2]   &                            &     \\
    \lstick{}                                 & \qw      & \targ                      & \qw                        & \qw & \ctrl{2} & \qw                   & \qw & \qw      & \meter & \cw                        & \push{\oplus} \cw \cwx[2]  & \cw \\
    \lstick{}                                 & \qw      & \qw                        & \targ                      & \qw & \qw      & \ctrl{2}              & \qw & \qw      & \meter & \push{\oplus} \cw \cwx[2]  & \cw                        & \cw \\
    \lstick{}                                 & \gate{H} & \ctrl{2}                   & \qw                        & \qw & \targ    & \qw                   & \qw & \qw      & \meter & \cw                        & \push{\oplus} \cw \cwx[2]  &     \\     
    \lstick{}                                 & \gate{H} & \qw                        & \ctrl{1}                   & \qw & \qw      & \targ                 & \qw & \qw      & \meter & \push{\oplus} \cw \cwx[1]  &                            &     \\
                                              &          &                            &                            &     &          &                       &     &          &        &                            &                            &     \\
                                              &          &                            &                            &     &          &    
                                              &     &          &        &                            &                            &     \\
    \hspace{-4em}\raisebox{1.5em}{$\vdots$} &          & \raisebox{1.5em}{$\vdots$} & \raisebox{1.5em}{$\vdots$} &     &          &                       &     &          &        & \raisebox{1.5em}{$\vdots$} & \raisebox{1.5em}{$\vdots$} &     \\
                                              &          & \qwx[1]                    & \qwx[2]                    &     &          &                       &     &          &        & \cwx[2]                    & \cwx[1]                    &     \\
    \lstick{}                                 & \qw      & \targ                      & \qw                        & \qw & \ctrl{2} & \qw                   & \qw & \qw      & \meter & \cw                        & \push{\oplus} \cw \cwx[2]  & \cw \\
    \lstick{}                                 & \qw      & \qw                        & \targ                      & \qw & \qw      & \ctrl{2}              & \qw & \qw      & \meter & \push{\oplus} \cw \cwx[2]  & \cw                        & \cw \\
    \lstick{}                                 & \qw      & \qw                        & \qw                        & \qw & \targ    & \qw                   & \qw & \ctrl{1} & \meter & \cw                        & \push{\oplus} \cw \cwx[1]  & \cw \\
    \lstick{}                                 & \qw      & \qw                        & \qw                        & \qw & \qw      & \targ                 & \qw & \targ    & \meter & \push{\oplus} \cw          & \push{\oplus} \cw          & \cw 
    \inputgroupv{1}{2}{0.8em}{0.8em}{\ket{00}_L\;\;}
    \inputgroupv{3}{4}{0.8em}{0.8em}{\ket{00}_L\;\;}
    \inputgroupv{5}{6}{0.8em}{0.8em}{\ket{00}_L\;\;}
    \inputgroupv{7}{8}{0.8em}{0.8em}{\ket{00}_L\;\;}
    \inputgroupv{13}{14}{0.8em}{0.8em}{\ket{00}_L\;\;}
    \inputgroupv{15}{16}{0.8em}{0.8em}{\ket{00}_L\;\;}
    \\
    }
    \]
    \caption{Arbitrary-length logical constant-depth CX ladder. A $\oplus$ on a measured data bit is flipped if the sum (modulo 2) of the measured ancilla bits above it with a $\oplus$ is 1. The qubits with/without a Hadamard applied at the start are ancilla/data qubits, respectively.}
    \label{fig:cdcx_arbitrary}
\end{figure}

\subsection{Shor Loss Correction Results}\label{sec:loss correction}
\begin{figure}
    \centering
    a.\\
    \includegraphics[width=.92\linewidth]{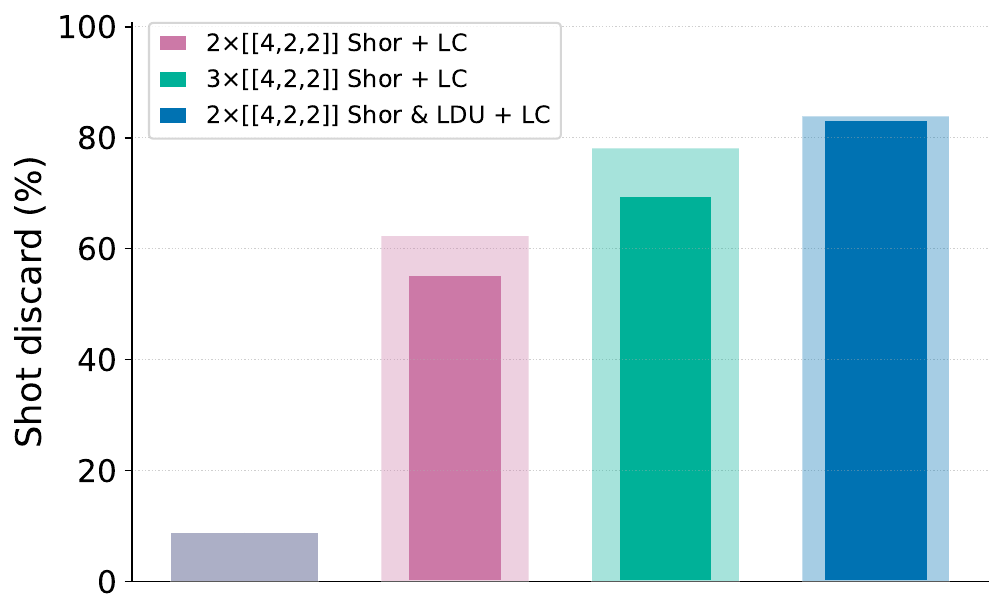}\\
    b.\\
    \includegraphics[width=.95\linewidth]{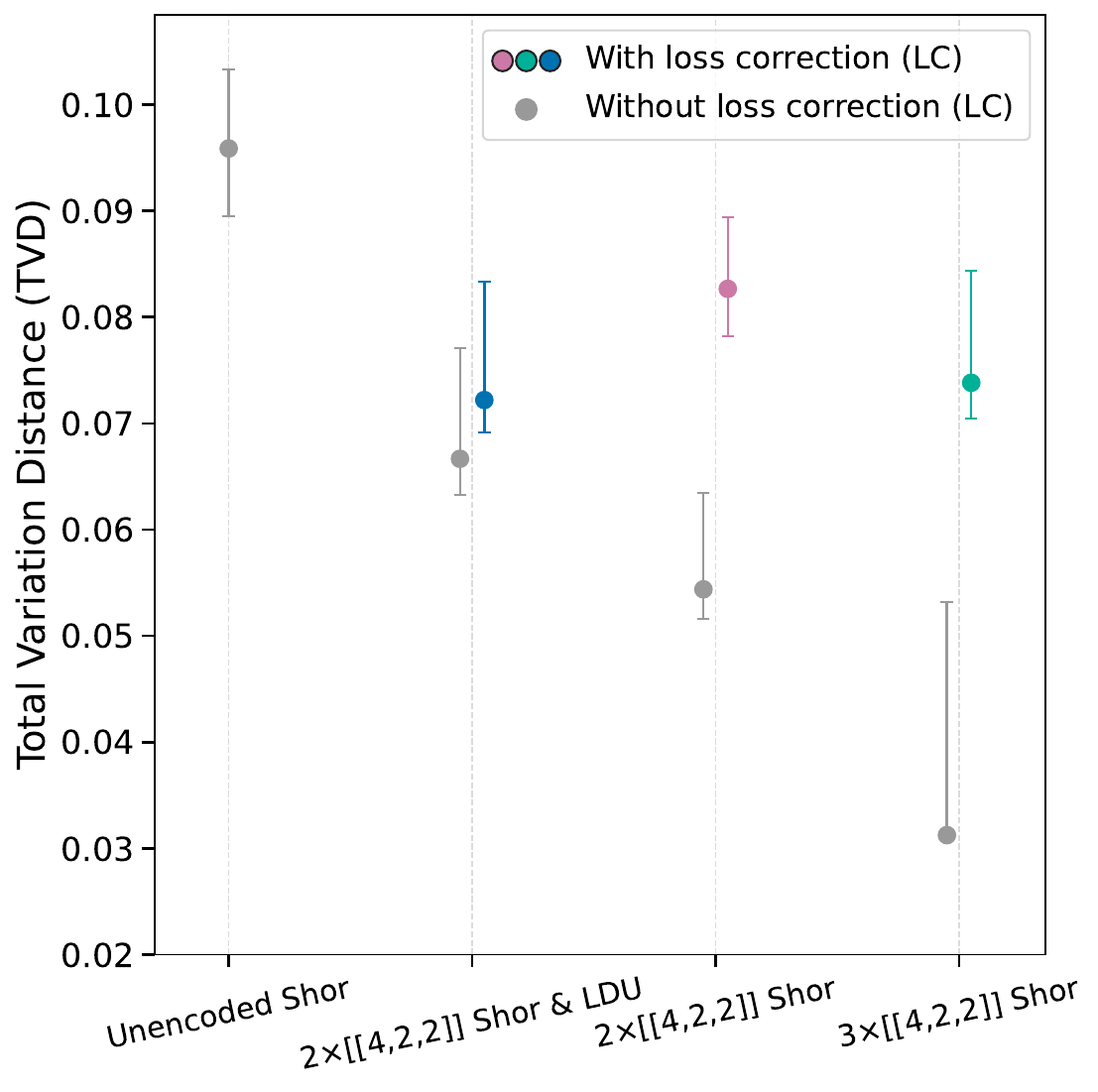}
    \caption{Updated experimental results for Shor circuits after loss correction (LC) post-processing.
    (a) The updated postselection discard where the thinner solid bars are loss corrected, while the translucent ones are not (that is, the same data from Fig. $\ref{fig:shor_experimental_data}$ for comparison), and
    (b) 
    The TVD computed between the experimental and ideal distributions with and without loss correction. Analysis the same as in Fig. $\ref{fig:shor_experimental_data}$.}
    \label{fig:shor_experimental_data_loss_corrected}
\end{figure}

As discussed in \cref{subsection:post-processing}, an additional tool we have in our logical encoding toolkit is that of loss correction. While the circuits executed in \cref{sec:shor} do not explicitly leverage a dedicated loss-correction gadget at the circuit-level, we can still perform loss correction by mapping each observed state to the closest  $[[4,2,2]]$ codeword if three of four qubits are retained. 
Fig. \ref{fig:shor_experimental_data_loss_corrected} depicts the experimental results from \cref{sec:shor} but with loss correction post-processing enabled for the logical encodings only. As expected, Fig. \ref{fig:shor_experimental_data_loss_corrected} shows that loss correction appreciably increases 
both the postselection yield and 
the TVD, although the latter continues to outperform the unencoded circuit. This is expected as the post-processing is not perfect. For example, shots including both a loss event and a bit-flip error on a different qubit can now be accepted, producing an erroneous result, when they would have been rejected without loss correction. We observe that the gap in performance between the two- and three-row Shor encodings narrows with loss correction, and we attribute this to the fact that there are simply more opportunities to infer an incorrect logical codeword with more [[4, 2, 2]] patches. This is consistent with the hypothesis that the superior performance of the three-row Shor encoding results from its additional error detection opportunities. Interestingly, this also has the effect of allowing the two-row Shor LDU circuit to maintain a lower TVD (than its non-LDU variant), as its additional pruning of leakage qubits will allow it to accept less of the potentially faulty loss corrected bitstring outcomes. This highlights a trade-off to consider: is the higher postselection yield preferable to a minor degradation in TVD (where the criterion for `minor’ is maintaining better than unencoded performance)? This trade-off may become more relevant in light of the CZ (or even movement) overhead of an actual loss-correction circuit gadget (similar to the LDU cost observed in the Shor simulations of \cref{sec:shor_noisy_simulations}) or in a shot-constrained scenario where low postselection yield becomes detrimental -- and further preferable if more sophisticated correction post-processing schemes are available to alleviate the apparent TVD (or similar performance metric) degradation.

\clearpage
\bibliography{refs}
\end{document}